\documentclass[aps, prd, twocolumn, showpacs, superscriptaddress, groupedaddress]{revtex4} 
\usepackage{graphicx}	
\usepackage{amssymb}
\usepackage{dcolumn}
\usepackage[mathscr]{euscript}
\usepackage{amsfonts}
\usepackage{amsmath}
\usepackage{color}
\usepackage{subfigure}
\usepackage{epsfig}
\usepackage{subfigure, rotating, bm, array}
\usepackage[pagebackref=false, colorlinks=true]{hyperref}
\hypersetup{
linkcolor=blue,     
citecolor=blue,     
urlcolor=blue}      
\begin{document}

\title{Tidal forces in collapsing compact objects}

\author{Ashok B. Joshi}
\email{gen.rel.joshi@gmail.com}
\affiliation{PDPIAS,
Charotar University of Science and Technology, Anand- 388421 (Guj), India.}
\author{Dipanjan Dey}
\email{deydipanjan7@gmail.com}
\affiliation{Department of Mathematics and Statistics, Dalhousie University, Halifax, Nova Scotia, Canada, B3H 3J5}
\author{Pankaj S. Joshi}
\email{psjcosmos@gmail.com}
\affiliation{International Centre for Space and Cosmology, School of Arts and Sciences, Ahmedabad University, Ahmedabad-380009 (Guj), India.}
\author{Vivekkumar R. Tank}
\email{vivek.tank.collapse@gmail.com}
\affiliation{PDPIAS,
Charotar University of Science and Technology, Anand- 388421 (Guj), India.}
\date{\today}

\begin{abstract}
In this work, we investigate tidal forces in the Lemaitre-Tolman-Bondi (LTB) metric, focusing on both hidden and locally visible singularities. We discuss the strength of these singularities in terms of deformationally strong singularities. Specifically, we analyze tidal forces in LTB spacetime, calculating radial and angular tidal forces and Jacobi fields for the radially co-moving shell. To provide a comparative study, we consider both homogeneous and inhomogeneous cases. The matter field distribution at one-time slice can differ significantly from another, highlighting the potential for time-dependent tidal deformation as a distinct observational signature. We focus on a specific feature: the time-varying maximum of stretching in the radial tidal force, which we term the "critical tidal boundary." In the inhomogeneous case, close to singularity time ($t<t_{s}$), the magnitudes of tidal forces vary substantially, with significant differences in compressive and stretching forces within a small physical radius $R(t,r)$. The resulting singularity in the LTB metrics at the end state of gravitational collapse appears to be an Ori-strong singularity, characterized by infinite tidal deformation.
\bigskip

$\boldsymbol{key words}$ : Naked singularity spacetime, Black hole spacetime.
\end{abstract}
\maketitle

\section{Introduction}
A massive object surpassing Chandrasekhar's limit will ultimately collapse into either a black hole or a neutron star. When a massive star exhausts its nuclear fuel, it undergoes a catastrophic collapse. This continuous gravitational collapse culminates in the formation of a spacetime singularity. Spacetime singularities are categorized into five distinct classes: future spacelike singularity, past spacelike singularity, future null singularity, past null singularity, and timelike singularity \cite{Joshi:2023ugm}. In the final stage of the gravitational collapse model proposed by Oppenheimer and Snyder, the singularity exists in the future of spacetime. The paper demonstrates that the endpoint of gravitational collapse is a hidden singularity, shrouded by the event horizon \cite{OppenheimerSnyder39}. The Lemaitre-Tolman-Bondi (LTB) model of gravitational collapse generalizes the dust solution to account for homogeneous and inhomogeneous matter distributions \cite{Joshi:1993zg, gcsc}.
As an object collapses, its center tends to densify progressively. Eventually, the core becomes so dense that it induces tidal disruption within the collapsing object. There are numerous potential scenarios leading to the formation of a spacetime singularity, but precise predictions about subsequent events remain challenging. The LTB metrics describe the spherically symmetric, pressure-less dynamical solutions of the Einstein field equation.

Theoretically, a black hole (BH) is not the only potential solution to Einstein's field equations. A good example is a naked singularity (NS) \cite{PhysRevLett.20.878,perlick,Joshi:2011zm, Joshi:2020tlq}. Under certain circumstances of gravitational collapse, null geodesics may escape from this kind of singularity (i.e., NS) and would be detectable to a distant observer \cite{Joshi:2011zm, Joshi:2013dva}. The appearance of the singularity might then vary depending on when the apparent horizon and trapped surfaces formed, appearing locally or globally. Alternative models of compact objects are suggested by recent observational investigations of modified gravity \cite{Khodadi:2022pqh, Khodadi:2021gbc, Khodadi:2020gns, KumarWalia:2022ddq}. The accretion disks \cite{Tahelyani:2022uxw, Patra:2023epx, Kovacs:2010xm}, dark matter \cite{Joshi:2022azj}, shadows cast by them \cite{Kumar:2020ltt, Shaikh:2022ivr, Ghosh:2021txu}, and particle trajectories around them \cite{Bambhaniya:2019pbr, Battista:2022krl} are the main subjects of research in the case of NS.

Tidal disruption is an extraordinary phenomenon observed in galaxies when a star approaches a supermassive black hole closely enough for its tidal forces to tear the star apart. This event generates electromagnetic radiation, providing insights into the properties of the corresponding supermassive black hole. Consequently, significant efforts have been dedicated to studying the effects of tidal forces, including the distortion of a body's shape within black hole spacetimes \cite{madan2022tidal}. In the context of Schwarzschild black hole spacetime, it has been established that tidal forces cause a body falling toward the black hole to elongate radially while compressing in the angular direction \cite{zhang2018tidal}.

Recent work on tidal forces in different types of spacetime geometries is discussed in \cite{Toshmatov:2023anz, Vandeev_2021, Arora:2023ijd, Arora:2023ltv}. When an object falls freely towards the spacetime center in Schwarzschild spacetime, it always experiences compression in the angular direction and stretching in the radial direction. Specifically, in the Reissner-Nordstrom case, the charge of the black hole significantly impacts the tidal force experienced by an object falling toward the center of the black hole. The radial and angular components of the tidal force can undergo a change of sign due to the presence of the charge \cite{Rntidal}. Other static spacetimes, such as those of regular black holes, Kiselev black holes, and naked singularities, have also been examined in relation to their tidal effects \cite{Li2017}. The impact of the Gauss-Bonnet coupling constant on tidal forces and the geodesic deviation vector has been analyzed in the four-dimensional Gauss-Bonnet black hole spacetime \cite{2021EPJC...81..590L}. Furthermore, investigations have shown that in the ergosphere of a rotating black hole, tidal interactions with a star can result in the emission of a jet comprised of debris. This phenomenon could potentially explain the formation of jets near black holes \cite{Li2017}. Consequently, the study of tidal effects in various types of black holes has become a topic of interest in the field of theoretical astrophysics.

The relative acceleration of two freely falling bodies can be used to determine the relative tidal force between two neighboring points. The geodesic deviation equation relates this relative acceleration directly to the curvature of spacetime. The geodesic deviation equation reads as:
\begin{equation}{\label{geodesic deviation homogeneous}}
    \frac{D^2\xi^\mu}{D\tau^2}-R^\mu_{\nu\rho\sigma}v^\nu v^\rho \xi^\sigma=0,
\end{equation}
where $\xi^{\mu}$ is the `Jacobi vector field' or \textit{Jacobi field} that represents the displacement to an infinitesimally nearby geodesic and is referred to as the deviation vector. $v^\nu$ is the tangent vector on the geodesic, $\gamma$ \cite{madan2022tidal}. If a Jacobi field $\xi^{\mu}$ is not identically zero but vanishes at both $p$ and $q \in \gamma$, then $p$ and $q$ are said to be conjugate. In other words, $p$ and $q$ are conjugate if a geodesic that is ``infinitesimally nearby" meets $\gamma$ at both $p$ and $q$. There need not be an actual geodesic other than $\gamma$ that runs through $p$ and $q$ for a Jacobi field to vanish at $p$ and $q$. A geodesic other than $\gamma$ passing between $p$ and $q$, on the other hand, does not imply that $p$ and $q$ are conjugate or even that a point conjugate to $p$ exists between $p$ and $q$. Conjugate points specify the location in spacetimes where a timelike geodesic fails to specify a local maximum of proper time between two points and a null geodesic fails to continue into the future of a point \cite{wald,hawking}.

To calculate tidal deformation on a particular co-moving radius within an Instantaneous Rest Frame (IRF), we define an orthonormal basis vector $\hat{e}_{\alpha}$. Due to spherical symmetry, points on constant radial surfaces exhibit the same tidal deformation property. The main purpose of this paper is to investigate tidal forces during gravitational collapse, with a focus on the significant information provided by tidal forces at the center of the collapsing body regarding the nature of singularities. From a physics perspective, our investigation helps to understand how the tidal effect on a freely falling object with negligible gravitational feedback varies over time as it approaches the center of a collapsing object. We show that tidal deformation can provide information regarding the causal structure of the singularity formed in a gravitational collapse.

The strength of singularities in LTB spacetime is largely discussed in terms of Tipler's strong singularity \cite{Tipler:1977zza,Clarke,Newman:1985gt,Waugh:1988ud}. One significant drawback with Tipler's original concept of strength is noted in Nolan's study \cite{Nolan:1999tw}. This definition is not based on the behavior of individual Jacobi fields but on the volume element formed by three of them. Examples where one Jacobi field shrinks to zero and another diverges can be created so that the volume element continues to be finite. In this instance, the singularity is characterized by Tipler's concept as weak even if a physical item will be torn apart (unboundedly stretched in one direction and compressed in another) \cite{Ori:2000fi}. Here, we investigate tidal forces in LTB spacetime and calculate the Jacobi field for the marginally bound LTB case. We consider two examples of strong singularities in Tipler's sense for homogeneous and inhomogeneous scenarios, focusing on tidal forces and the Jacobi field.

The plan of the paper is as follows. In section (\ref{sec1}), we discuss the dynamics of collapsing spacetime in Lemaitre-Tolman-Bondi (LTB) spacetime. We discuss matching conditions and conditions for the visibility of singularity. In section (\ref{sec2}), we discuss tidal forces in gravitational collapse for homogeneous and inhomogeneous dust collapse. We also calculate the Jacobi field for the marginally bound cases. We also provide a comparative study of the tidal force effect in homogeneous and inhomogeneous dust collapse. Finally, our results are discussed in Section (\ref{sec3}), followed by the conclusion. Throughout the paper, we take Newton's gravitational constant ($G$) and the speed of light ($c$) as unity and the metric signature $(-,+,+,+)$.

\section{Lemaitre-Tolman-Bondi metric}\label{sec1}
An astronomical object undergoes gravitational collapse due to the contraction caused by its self-gravity. This scenario can be modeled using a dynamic spacetime described by the spherically symmetric dust solution of Einstein's field equations, commonly referred to as the Lemaitre-Tolman-Bondi (LTB) metric. The line element of the LTB metric is:
\begin{equation}
 ds^2 =  -dt^2 + \frac{R'^2(t,r)}{G(r)}dr^2 + R^2(t,r)d\Omega^2\, , \label{Metric}
\end{equation}
where $R(t,r)$ represents the physical radius, with $(t,r)$ denoting comoving time and radius. Additionally, we define a positive real-valued function $G(r)=1+f(r)$, where $f(r)$ is the velocity function. For the marginally bound case under consideration, $f(r)=0$ \cite{gcsc}. Consequently, equation \eqref{Metric} simplifies to:
\begin{equation}
ds^2= -dt^2 + R'^2(t,r)dr^2 + R^2(t,r)d\Omega^2\, . \label{spacetime}
\end{equation}
where $d\Omega^2 = d\theta^{2} + \sin^{2}\theta \, d\phi^{2}$.

From Einstein's field equations, one can obtain the density and pressure as follows:
\begin{equation}
    \rho=\frac{F'}{R^2 R'}, \quad P_r = -\frac{\dot{F}}{R^2 \dot{R}}
    \label{Rho}
\end{equation}
\begin{equation}
    P_\perp = P_\theta = P_\phi = P_{r} + \frac{P_{r}' R}{2 R'}, \quad F(t,r) = R \dot{R}^2.
    \label{tanpressure}
\end{equation}
Solving the right side of equation \eqref{tanpressure}, we find:
\begin{equation}
    R(t,r) = \left(r^\frac{3}{2} -\frac{3}{2} \sqrt{F(r)} t \right)^\frac{2}{3}. \label{physicalR}
\end{equation}
In equations \eqref{Rho} and \eqref{tanpressure}, $\rho$, $P_r$, and $P_\perp$ denote the energy density, radial pressure, and azimuthal (tangential) pressure, respectively.

As mentioned before, the fluid that seeds LTB spacetime is dust. Therefore, to ensure the satisfaction of the strong, weak, and null energy conditions, it can be readily verified that we need $R' >0$ when $F' >0$, where $R' >0$ also ensures the absence of any shell-crossing singularity.  

\subsection{Matching condition and outside metric}
A smooth connection between two spacetimes at a timelike or spacelike hypersurface requires satisfying two junction conditions on the corresponding hypersurface \cite{refn1}. First, both sides of the matching hypersurface ($\Sigma$) must have the same induced metric. Second, in both the internal and external spacetimes, the extrinsic curvature of the matching hypersurface needs to match. The extrinsic curvature is expressed as: 
\begin{equation}
    K_{ab} = e^{\alpha}_{a}e^{\beta}_{b}\nabla_{\alpha}\eta_{\beta},
\end{equation}
where $\eta^{\beta}$ and $e^{\alpha}_{a}$ are the normal and tangents to the hypersurfaces, respectively. 

For the generic cases of gluing two Lorentzian manifolds, $\mathcal{M}_{int}$ (collapsing body) and $\mathcal{M}_{ext}$ (exterior metric), with a smooth continuous manifold, both conditions are necessary and sufficient. The process involves joining spacetimes via thin shell formalism across a boundary hypersurface denoted as $\Sigma = \partial \mathcal{M}_{int} \cap \partial\mathcal{M}_{ext}$. The total Lorentzian manifold could be $\mathcal{M} = \mathcal{M}_{int} \cup \mathcal{M}_{ext}$ \cite{Israel:1966rt}. In the LTB case, the outside matching metric is considered to be a Schwarzschild metric, which is a vacuum solution of the Einstein field equation. Thus, we can write the external Schwarzschild metric as:
\begin{equation}
    ds^2 = -\left(1 - \frac{2M_{ADM}}{\mathcal{R}(\nu)} + 2\frac{d\mathcal{R}(\nu)}{d\nu}\right)d\nu^2 + \mathcal{R}(\nu)^2 d\Omega^2 \label{schmatch}
\end{equation}
Here, $\nu$ is the retarded null coordinate, $M_{ADM}$ is the ADM mass of spacetime, and $\mathcal{R}(\nu)$ is the evolving radius. 

Matching Eq.~\eqref{schmatch} with Eq.~\eqref{Metric} at the hypersurface $r = R_{b}$, for the zero radiation case, we obtain:
\begin{equation}
    \mathcal{R}(\nu) = R(t, R_{b}) = R_{b} \, a(t),
\end{equation}
\begin{equation}
    F(R_{b}) = 2M_{ADM}, \label{adm}
\end{equation}
\begin{equation}
    \ddot{\mathcal{R}}(\nu) = -\frac{F(R_{b})}{2\mathcal{R}(\nu)^2} = -\frac{M_{ADM}}{\mathcal{R}(\nu)^2}.
\end{equation}
For the homogeneous case, Eq.~\eqref{adm} gives $F_{0} = (2M_{ADM})/R_{b}^{3}$. Two factors can be freely chosen in this case: the collapsing body's mass and its boundary, which is its outermost co-moving shell. Matching can yield the values of $F_{0}$ and $F_{3}$ for the mass function $F(r) = F_{0}r^{3} - F_{3}r^{6}$ in the inhomogeneous case as follows:
\begin{equation}
    F_{0} = \frac{4M_{ADM}}{R_{b}^3} - \frac{1}{3}\rho(0,R_{b}) \label{f0}
\end{equation}
\begin{equation}
     F_{3} = \frac{2M_{ADM}}{R_{b}^6} - \frac{\rho(0,R_{b})}{3R_{b}^3}.\label{f3}
\end{equation}
Since the mass function is positive, it suggests that $F_{0} > 0$ and $F_{3} > 0$. Our initial density profile for the given mass function can be expressed as:
\begin{equation}
    \rho(r) = 3F_{0} - 6F_{3}r^{3}.\label{densityp}
\end{equation}
Substituting Eq.~\eqref{f0} and Eq.~\eqref{f3} into Eq.~\eqref{densityp}, we obtain:
\begin{equation}
    \rho(r) = \frac{12M_{ADM}}{R_{b}^{3}} \left(1-\frac{r^{3}}{R_{b}^{3}}\right) - \rho(0,R_{b})\left(1-\frac{2r^{3}}{R_{b}^{3}}\right) .\label{density}
\end{equation}
Generally, the density should be a monotonically decreasing function, where the core is denser than the boundary surface. Hence, Eq.~\eqref{densityp} suggests that $F_{3}$ should be positive. This leads us to the conclusion that the initial density at the boundary radius must satisfy the requirement $\rho(0, R_{b}) < 6M_{ADM}/R_{b}^{3}$ \cite{Joshi:2024kqd}.

\subsection{Visibility of singularity}
To demonstrate that a singularity is (at least locally) naked, it is necessary to show the existence of non-spacelike future-directed outgoing geodesics with their past endpoint at the singularity.
As discussed in \cite{Joshi:2023ugm}, for $F(r) = F_{0}r^{3} - F_{3}r^{6}$, setting $F_{0} = 1$ results in a locally visible singularity when $F_3 > \sim 25.967$. For the above mass function, the singularity is either locally or globally visible depending on the choice of the coefficients $F_{0}$ and $F_{3}$. Using Eq.~\eqref{f0} and Eq.~\eqref{f3}, the condition for local visibility $F_{3} > 25.967$ yields:
\begin{equation}
    M_{ADM} = \frac{R_{b}^{3}}{4}, \hspace{1cm} R_{b} < 0.268\, . \label{conditions}
\end{equation}
For a locally visible singularity, the initial density at $r = 0$ (from Eq.~\eqref{density}),
\begin{equation}
    \rho(0) = \frac{12M_{ADM}}{R_{b}^{3}} - \rho(0, R_{b}),
\end{equation}
where $\rho(0, R_{b}) = 0$.

\begin{figure*}[ht!]
    \centering
\subfigure[$t=0.0\;$]
{\includegraphics[width=5.5cm]{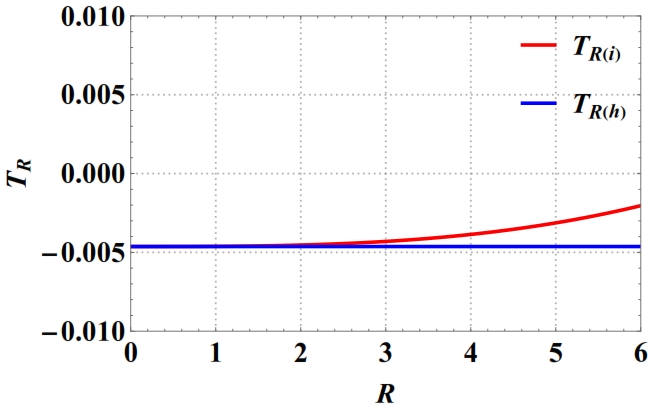}\label{fig:t=0}}
\hspace{0.5cm}
\subfigure[$t=1.0\;$]
{\includegraphics[width=5.5cm]{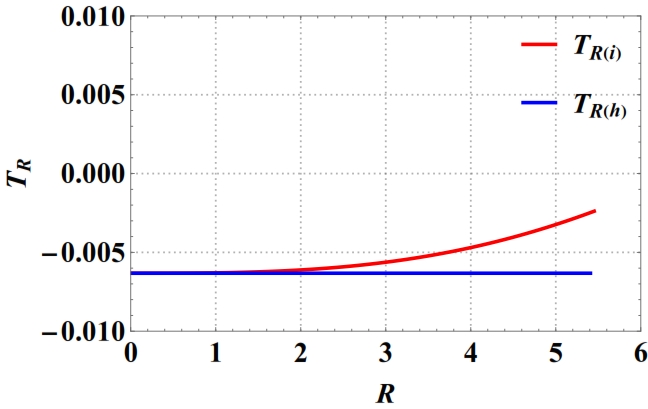}\label{fig:t=1}}
\hspace{0.5cm}
\subfigure[$t=2.0\;$]
{\includegraphics[width=5.5cm]{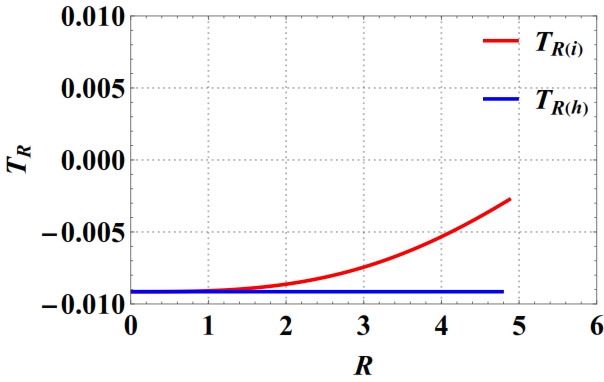}\label{fig:t=2}}\\
\subfigure[$t=3.0\;$]
{\includegraphics[width=5.5cm]{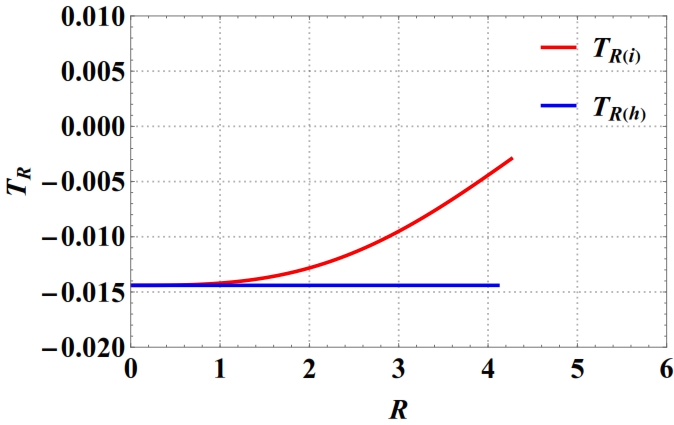}\label{fig:t=3}}
\hspace{0.5cm}
\subfigure[$t=4.0\;$]
{\includegraphics[width=5.5cm]{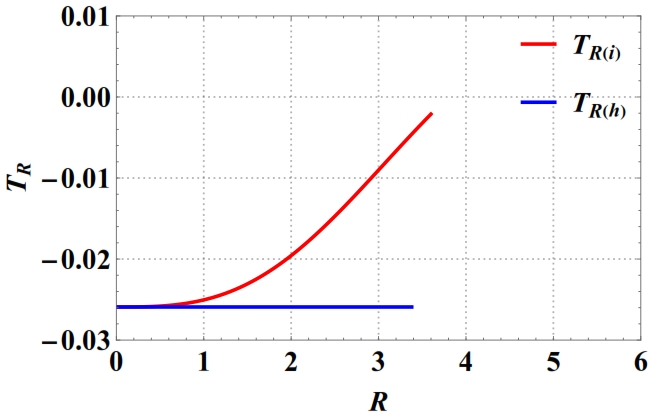}\label{fig:t=4 sec}}
\hspace{0.5cm}
\subfigure[$t=5.0\;$]
{\includegraphics[width=5.5cm]{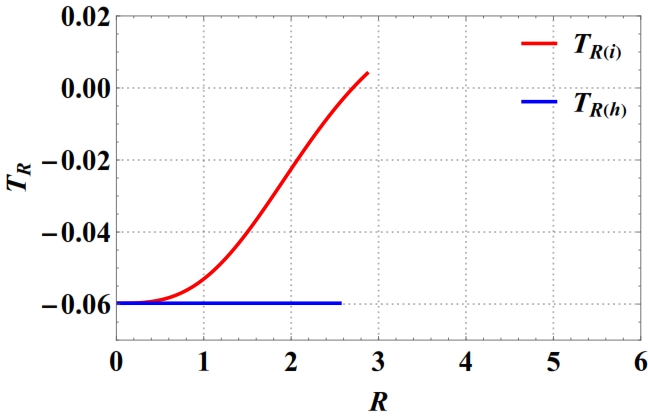}\label{fig:t=5 sec}}\\
\subfigure[$t=6.0\;$]
{\includegraphics[width=5.5cm]{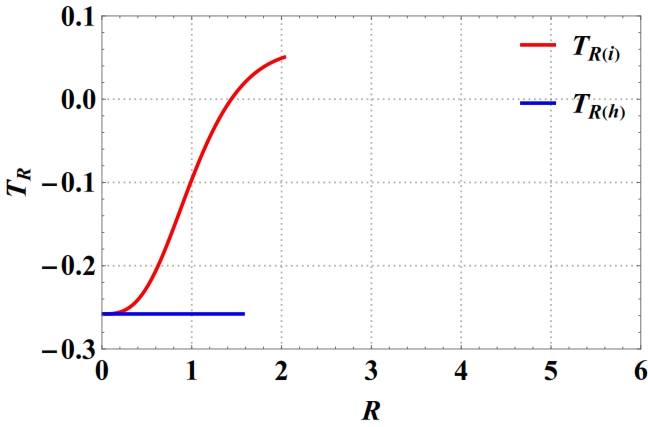}\label{fig:t=6}}
\hspace{0.5cm}
\subfigure[$t=6.5\;$]
{\includegraphics[width=5.5cm]{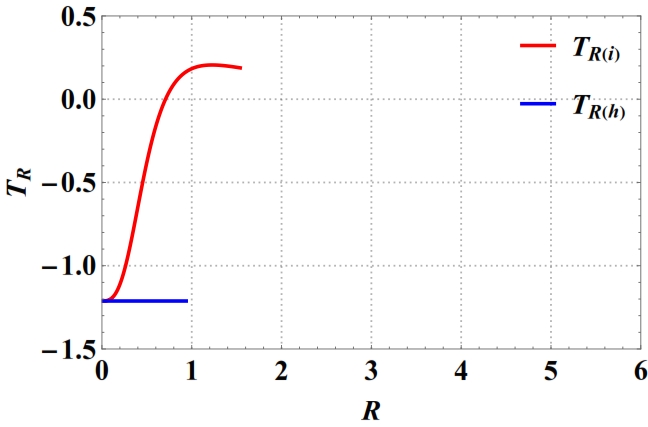}\label{fig:t=6.5}}
\hspace{0.5cm}
\subfigure[$t=6.9$]
{\includegraphics[width=5.5cm]{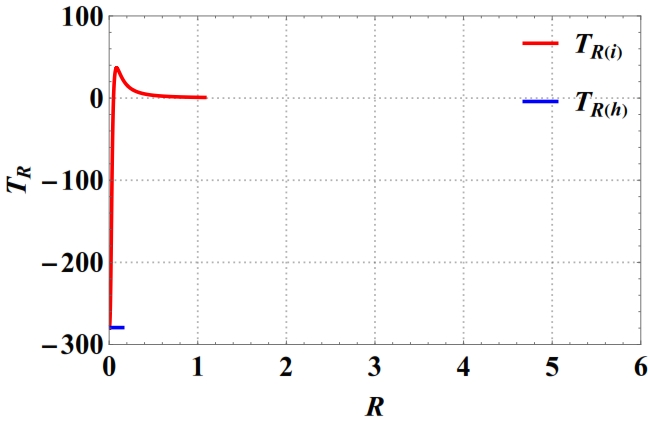}\label{fig:t=6.9}}\\
\subfigure[$t=t_s=6.92$]
{\includegraphics[width=5.5cm]{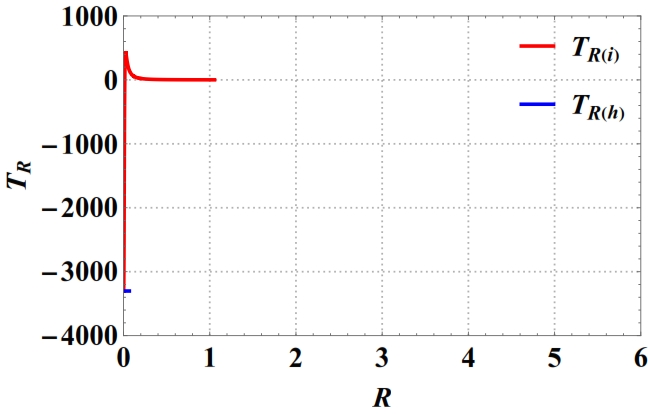}\label{fig:t=6.92x}}
\hspace{0.5cm}
\subfigure[$t=7.26$]
{\includegraphics[width=5.5cm]{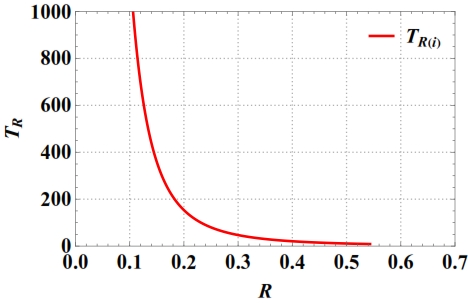}\label{fig:t=7.26 sec}}
\hspace{0.5cm}
\subfigure[$t=7.35$]
{\includegraphics[width=5.5cm]{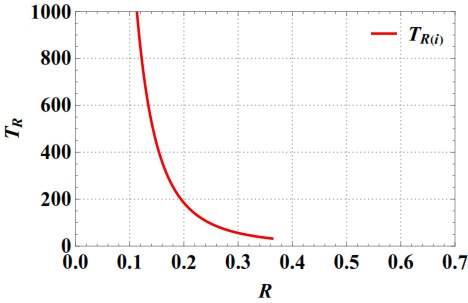}\label{fig:t=7.35}}
 \caption{This figure shows the radial tidal force ($T_R$) versus physical radius $(R)$ at different time slices for both homogeneous and inhomogeneous dust collapse. Here, $T_{R(i)}$ is the radial tidal force for the inhomogeneous case, and $T_{R(h)}$ is the radial tidal force for the homogeneous case. Here we take $M_{ADM} = 1$, $R_{b} = 6M_{ADM}$ and $\rho(R_{b}) = 0$.}
 \label{fig:1}
\end{figure*}

\begin{figure*}[ht!]
\centering
\subfigure[$t=0.0$]
{\includegraphics[width=5.5cm]{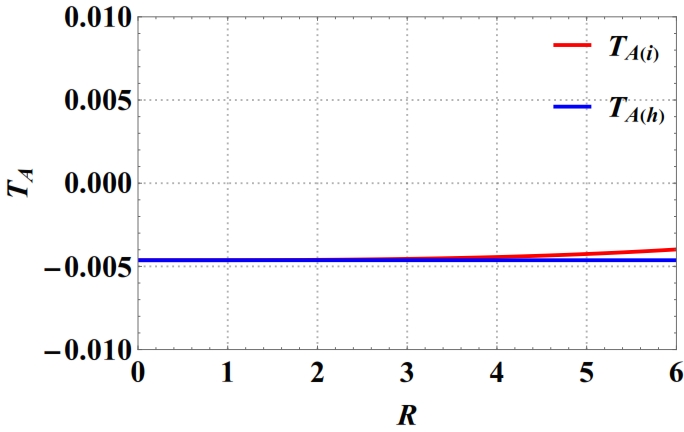}\label{fig:t=0.0}}
\hspace{0.3cm}
\subfigure[$t=1.0$]
{\includegraphics[width=5.5cm]{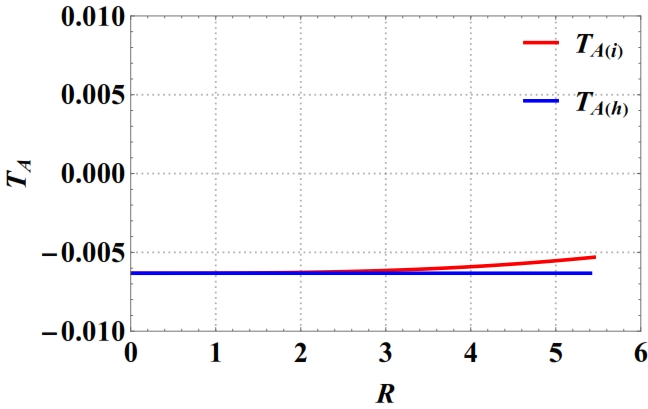}\label{fig:t=1.0}}
\hspace{0.3cm}
\subfigure[$t=2.0$]
{\includegraphics[width=5.5cm]{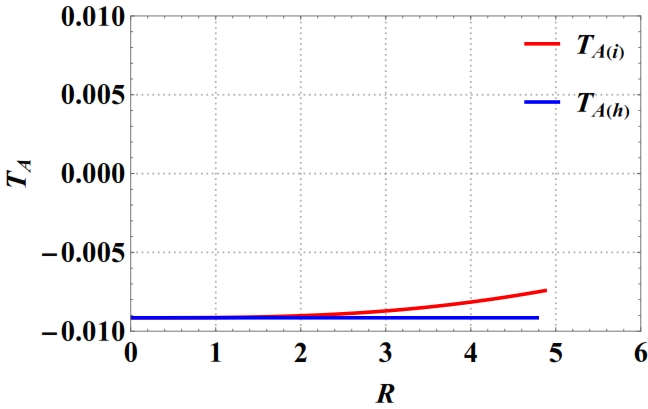}\label{fig:t=2.0}}\\
\subfigure[$t=3.0$]
{\includegraphics[width=5.5cm]{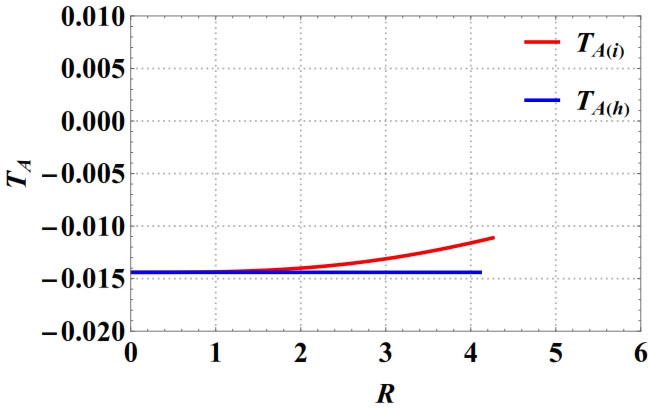}\label{fig:t=3.0}}
\hspace{0.3cm}
\subfigure[$t=4.0$]
{\includegraphics[width=5.5cm]{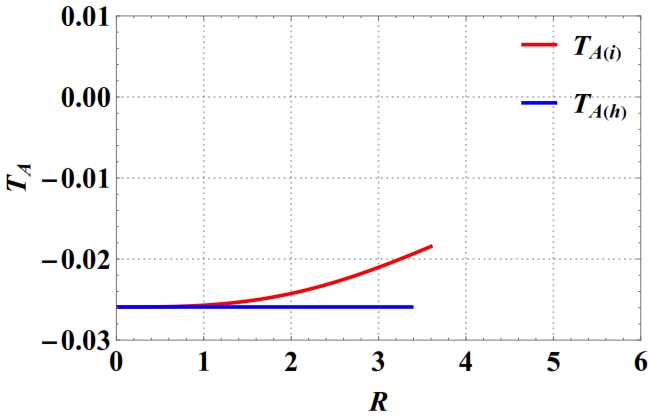}\label{fig:t=4.0}}
\hspace{0.3cm}
\subfigure[$t=5.0$]
{\includegraphics[width=5.5cm]{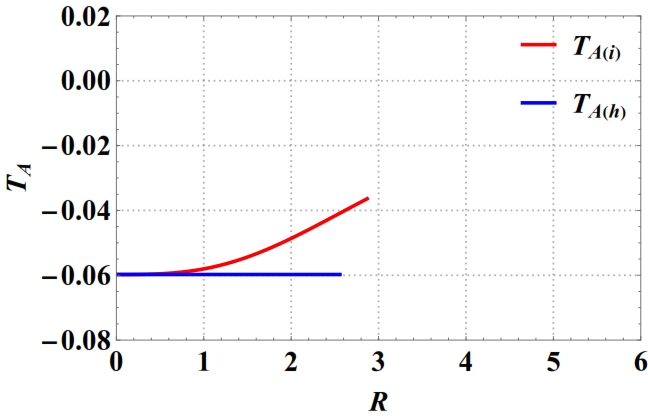}\label{fig:t=5.0}}\\
\subfigure[$t=6.0$]
{\includegraphics[width=5.5cm]{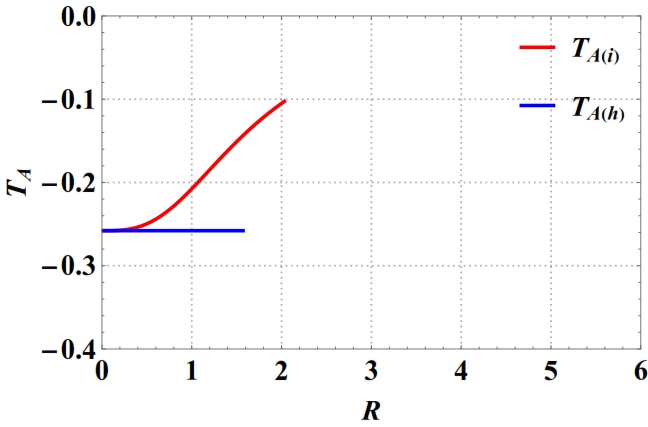}\label{fig:t=6.0}}
\hspace{0.3cm}
\subfigure[$t=6.5$]
{\includegraphics[width=5.5cm]{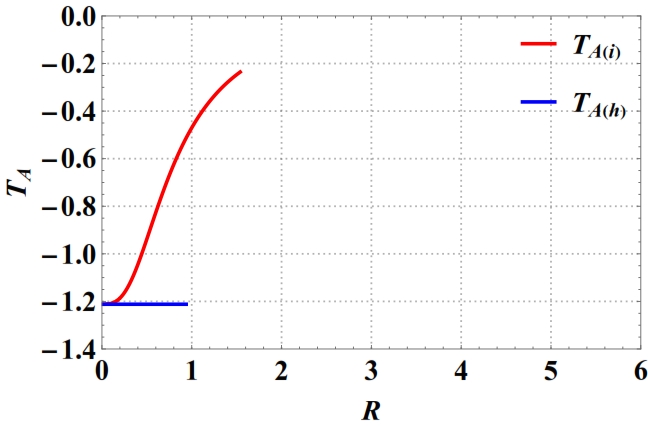}\label{fig:t=6.50}}
\hspace{0.3cm}
\subfigure[$t=6.9$]
{\includegraphics[width=5.5cm]{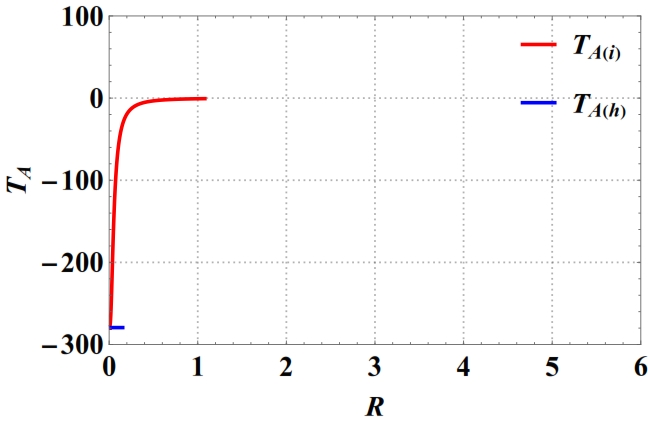}\label{fig:t=6.90}}\\
\subfigure[$t=t_s=6.92$]
{\includegraphics[width=5.5cm]{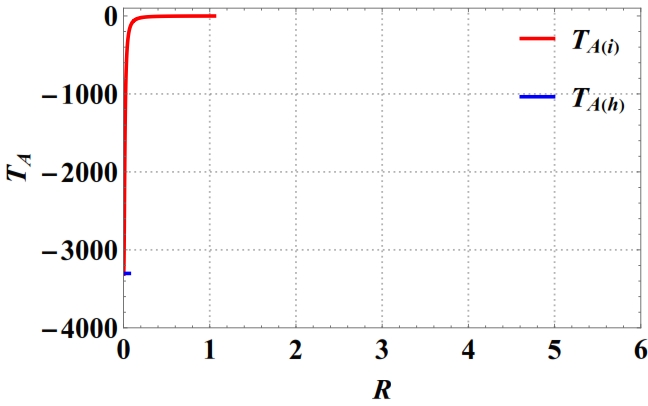}\label{fig:t=6.92}}
\hspace{0.3cm}
\subfigure[$t=7.26$]
{\includegraphics[width=5.5cm]{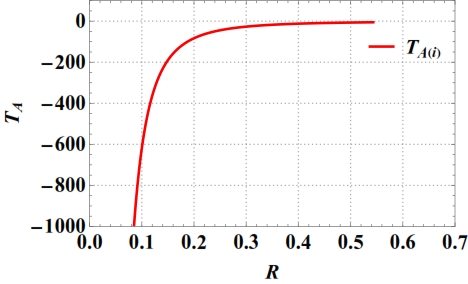}\label{fig:t=7.26}}
\hspace{0.3cm}
\subfigure[$t=7.35$]
{\includegraphics[width=5.5cm]{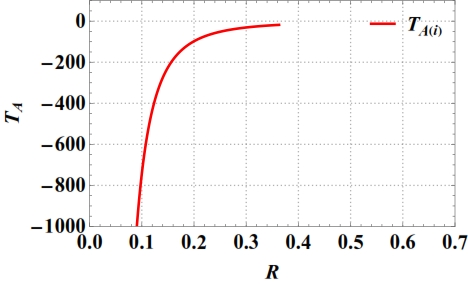}\label{fig:t=7.35}}
 \caption{This figure shows the angular tidal force ($T_A$) versus physical radius $(R)$ at different time slices for both homogeneous and inhomogeneous dust collapse. Here, $T_{A(i)}$ is the angular tidal force for the inhomogeneous case, and $T_{A(h)}$ is the angular tidal force for the homogeneous case. Here we take $M_{ADM} = 1$, $R_{b} = 6M_{ADM}$ and $\rho(R_{b}) = 0$.}
 \label{fig:2}
\end{figure*}

\section{Tidal force in the gravitational collapse}{\label{sec2}}
In this section, we investigate the effect of tidal forces on a self-gravitating object, particularly within the Lemaître-Tolman-Bondi (LTB) spacetime. To explore tidal forces in the framework of general relativity, we analyze the geodesic deviation equation:

\begin{equation}
    \frac{D^2 \eta^\mu}{D\tau^2} - R^\mu_{\nu \rho \sigma} v^\nu v^\rho \eta^\sigma = 0, \label{deviationeql}
\end{equation}
where $ R^\mu_{\nu \rho \sigma} $ and $ v^\nu $ denote the Riemann curvature tensor and the unit tangent vector of the geodesic, respectively, and $ \eta^\mu $ represents the geodesic deviation vector. In the presence of strong energy conditions and non-zero curvature, each point on the test body follows a unique geodesic, leading to stretching and/or squeezing—known as tidal effects. To quantify these effects, we employ the Jacobi field, which measures the distance between infinitesimally close geodesics.

For the LTB metric given in Eq. (\ref{Metric}), the tetrad basis related to the freely falling frame is expressed as:
\begin{eqnarray}
\hat{e}^\mu_{\hat{0}} &=& \{1, 0, 0, 0\}, \nonumber \\
\hat{e}^\mu_{\hat{1}} &=& \left\{0, \frac{\sqrt{1+f(r)}}{R'(t,r)}, 0, 0\right\}, \nonumber \\
\hat{e}^\mu_{\hat{2}} &=& \{0, 0, \frac{1}{R(t,r)}, 0\}, \nonumber \\
\hat{e}^\mu_{\hat{3}} &=& \{0, 0, 0, \frac{1}{R(t,r)\sin\theta}\}. \label{tetrad}
\end{eqnarray}
These satisfy the orthonormality condition:
\begin{equation}
    \hat{e}_{\alpha \hat{\mu}} \cdot \hat{e}^\alpha_{\hat{\nu}} = \eta_{\hat{\mu}\hat{\nu}},
\end{equation}
where $ \eta_{\hat{\mu}\hat{\nu}} = \text{diag}(-1, 1, 1, 1) $ represents components of the Minkowski metric. The hat indices denote the tetrad basis, while those without denote the coordinate basis. Additionally, $ \hat{e}^\mu_0 = v^\mu $, implying that the tetrad basis equals the 4-velocity vector of the observer when timelike \cite{zhang2018tidal}. The unit vectors $ \{\hat{e}^\mu_1, \hat{e}^\mu_2, \hat{e}^\mu_3\} $ correspond to orthogonal spatial directions in the observer's frame \cite{chandrasekhar1983mathematical}. Using the tetrad basis from Eq. (\ref{tetrad}), the geodesic deviation vector (or separation vector) can be expanded as:
\begin{equation}
    \xi^\mu = \hat{e}^\mu_{\hat{\nu}} \xi^{\hat{\nu}}.
\end{equation}
It's important to note that for a fixed temporal component, $ \xi^{\hat{0}} = 0 $ \cite{madan2022tidal}.
Next, we calculate the Riemann curvature tensor with respect to the tetrad basis using the tetrad formalism, given by:

\begin{equation}{\label{riemann tensor}}
    R^{\hat{a}}_{\hat{b}\hat{c}\hat{d}} = R^\mu_{\nu\rho\sigma} \hat{e}^{\hat{a}}_\mu \hat{e}^\nu_{\hat{b}} \hat{e}^\rho_{\hat{c}} \hat{e}^\sigma_{\hat{d}}.
\end{equation}

In the instantaneous rest frame (IFR), Eq.~(\ref{deviationeql}), can be expressed as
\begin{equation}
\frac{d^2\,\eta^{\hat{\alpha}}}{d\tau^2} = R^{\hat{\alpha}}_{ \hat{0} 
\hat{0} \hat{\gamma}} \, \eta^{\hat{\gamma}},\label{Rhat}
\end{equation}
where, 
$R^{\hat{\alpha}}_{ \hat{\beta} \hat{\gamma} 
\hat{\delta}}=R^a_{bcd}e^{\hat{\alpha}}_a e_{\hat{\beta}}^b 
e_{\hat{\gamma}}^c e_{\hat{\delta}}^d$.
Considering the vectors are parallelly transported along the 
geodesic and exploring the above equations (\ref{tetrad}) and 
(\ref{Rhat}), we obtain the relative acceleration between two nearby particles in radial and tangential directions as follows:
\begin{equation}
\frac{D^2 \eta^{\hat{r}}}{D\tau^2}= \frac{\Ddot{R}^{'}(t,r)}{R^{'}(t,r)} \eta^{\hat{r}},\qquad
\frac{D^2 \eta^{\hat{i}}}{D\tau^2}= \frac{\Ddot{R}(t,r)}{R(t,r)} \eta^{\hat{i}}, \label{tidalg}
\end{equation}
where $i = \theta, \phi$. Interestingly, the radial and angular components of tidal force vanish when $\Ddot{R}^{'}(t,r) = 0$ and $\Ddot{R}(t,r) = 0$, respectively. The above two equations represent the tidal force for a radially free-falling frame that is, an observer is comoving with collapsing fluid in the LTB spacetime. Hence, comoving time is the proper time in the fluid's frame, $\tau = t$. Therefore using Eq.~(\ref{physicalR}) and Eq.~(\ref{tidalg}) we can write, 
\begin{eqnarray}
    \frac{d^2 \eta^{\hat{r}}}{dt^2} = \frac{2 \sqrt{F(r)} \left(4 \sqrt{r} F(r) - F'(r) P(t,r)\right)}{Q(t,r)^2 \left(2 \sqrt{r} \sqrt{F(r)} - t F'(r)\right)} \eta^{\hat{r}}
\end{eqnarray}
\begin{eqnarray}
    \frac{d^2 \eta^{\hat{i}}}{dt^2} = - \frac{2 F(r)}{Q(t,r)^2} \eta^{\hat{i}}
\end{eqnarray}
Where, 
\begin{eqnarray}
    P(t,r) &=& \left(2 r^{3/2} - t \sqrt{F(r)}\right),\\
    Q(t,r) &=& \left(2 r^{3/2} - 3 t \sqrt{F(r)}\right).
\end{eqnarray}
Any smooth vector field along a causal geodesic that satisfies Eq.~(\ref{deviationeql}) is called a Jacobi field $\eta$. Since Eq.~(\ref{deviationeql}) is a second-order differential equation that depends on the initial values of $\eta^{\hat{\alpha}}$ and $\dot{\eta}^{\hat{\alpha}}$ at an initial time $t=0$, there are six independent Jacobi fields. Now solving the second order differential equation we get,

\begin{widetext}
    \begin{equation}
       \eta^{\hat{r}} = \frac{\left(2 r^{1/6} F'(r)^{2/3} A(t,r) \Delta - 2^{5/3} \sqrt{r} C_{2} F(r) \Sigma + \sqrt{F(r)} F'(r) \left(-4 r^{2/3} C_{1} F'(r)^{2/3} A(t,r) + 2^{2/3} t C_{2} \Sigma \right)\right)}{\left(2 r^{1/6} \sqrt{B(t,r)} F'(r)^{5/3} A(t,r)\right)}\label{etar},
    \end{equation}
    \begin{equation}
        \eta^{\hat{i}} = \left(C_{1} + C_{2} \left(2 r^{3/2} - 3 t \sqrt{F(r)}\right)^{1/3}\right) \left(2 r^{3/2} - 3 t \sqrt{F(r)}\right)^{1/3} ,
    \end{equation}
    \begin{equation}
         X(t,r) = 2 \sqrt{3} \arctan{\left(-\frac{1}{\sqrt{3}} + \frac{\left(2^{2/3} \sqrt{B(t,r)} F'(r)^{1/3}\right)}{\sqrt{3} r^{1/6} A(t,r)}\right)} ,
    \end{equation}
    \begin{equation}
         Y(t,r) = 2 \log{ \left(2^{2/3} \sqrt{B(t,r)} F'(r)^{1/3} + 2 r^{1/6} A(t,r)\right)} ,
    \end{equation}
    \begin{equation}
         Z(t,r) = \log{ \left(2^{1/3} B(t,r) F'(r)^{2/3} - 2^{2/3} r^{1/6} \sqrt{B(t,r)}\\ F'(r)^{1/3} A(t,r) + 2 r^{1/3} A(t,r)^2 \right)} ,
    \end{equation}
\end{widetext}
where, $C_{1}$ and $C_{2}$ are integration constant and,
    \begin{eqnarray}
        \Delta &=& \left(- C_{2} B(t,r) + t C_{1} F'(r)^2 \right), \nonumber\\
        \Sigma &=& \left( X(t,r) - Y(t,r) + Z(t,r)\right), \nonumber\\
        A(t,r) &=& \left(-3 F(r) + r F'(r)\right)^{1/3}, \nonumber\\
        B(t,r) &=& \left(-2 r^{3/2} + 3 t \sqrt{F(r)}\right)^{2/3}. \nonumber
    \end{eqnarray}
The extremum of the radial tidal force function can be calculated as follows: 
\begin{equation}
    \frac{d}{dR}\left(\frac{\Ddot{R}^{'}(t,r)}{R^{'}(t,r)}\right) = 0.\label{tidalhorizon}
\end{equation}
The maximum positive value of radial tidal force at a constant time slice and fixed physical radius provides information about the maximum stretching within the collapsing body. The time-evolving maxima of radial tidal force develop a critical boundary where the fluid properties differ significantly. For ease of reference, we term it the critical tidal boundary.

\subsection{Tidal force effect in homogeneous dust collapse}
For the homogeneous dust collapse case, we adopt the Misner-Sharp mass function as $F(r) = F_{0}r^{3}$. Then, the LTB metric for marginally bound homogeneous dust collapse is given by,
\begin{equation}\label{LTB metric homogeneous}
\begin{split}
    ds^2 = -dt^2 + \left(1 - \frac{3}{2}\sqrt{F_0} t\right)^{4/3} dr^2 \\
          + r^2 \left(1 - \frac{3}{2}\sqrt{F_0} t\right)^{4/3} d\theta^2 \\
          + r^2 \sin^2\theta \left(1 - \frac{3}{2}\sqrt{F_0} t\right)^{4/3} d\phi^2
\end{split}
\end{equation}
Our research delves into the intricate tidal force equations embedded within the fabric of spacetime, with particular emphasis on the LTB metric for homogeneous dust collapse. In this spacetime, the tetrad basis related to the freely falling frame can be expressed as \cite{Li2017}:
\begin{equation}\label{tetrad basis}
\begin{aligned}
\hat{e}^\mu_0 &= \left\{1, 0, 0, 0\right\}, \\
\hat{e}^\mu_1 &= \left\{0, \frac{1}{\left(1 - \frac{3}{2}\sqrt{F_0} t\right)^{2/3}}, 0, 0\right\}, \\
\hat{e}^\mu_2 &= \left\{0, 0, \frac{1}{r \left(1 - \frac{3}{2}\sqrt{F_0} t\right)^{2/3}}, 0\right\}, \\
\hat{e}^\mu_3 &= \left\{0, 0, 0, \frac{1}{r \left(1 - \frac{3}{2}\sqrt{F_0} t\right)^{2/3} \sin\theta}\right\}.
\end{aligned}
\end{equation}
Using the expression from Eq. (\ref{riemann tensor}), we calculate the Riemann curvature tensor in terms of the tetrad basis:
\begin{equation}\label{radial component of tidal force in homogeneous dust collapse}
    R^{\hat{1}}_{\hat{0}\hat{0}\hat{1}} = -\frac{2F_{0}}{\left(2 - 3\sqrt{F_{0}} t\right)^2},
\end{equation}
\begin{equation}\label{angular component of tidal force in homogenenous dust collapse}
    R^{\hat{2}}_{\hat{0}\hat{0}\hat{2}} = R^{\hat{3}}_{\hat{0}\hat{0}\hat{3}} = -\frac{2F_{0}}{\left(2 - 3\sqrt{F_{0}} t\right)^2}.
\end{equation}
Here, in Eq. (\ref{radial component of tidal force in homogeneous dust collapse}), $R^{\hat{1}}_{\hat{0}\hat{0}\hat{1}}$ represents the radial component of the tidal force, and in Eq. (\ref{angular component of tidal force in homogenenous dust collapse}), $R^{\hat{2}}_{\hat{0}\hat{0}\hat{2}}, R^{\hat{3}}_{\hat{0}\hat{0}\hat{3}}$ represent the angular components of the tidal force in the LTB metric for homogeneous dust collapse \cite{zhang2018tidal}. By substituting these values of the Riemann tensors into the geodesic deviation equation (\ref{geodesic deviation homogeneous}), we obtain the equations for the tidal force in a radially freely falling reference frame:
\begin{equation}
    \frac{D^2 \xi^{\hat{1}}}{D\tau^2} = \Ddot{\xi}^{\hat{1}} = -\frac{2F_{0}}{\left(2 - 3\sqrt{F_{0}} t\right)^2} \xi^{\hat{1}}, \label{EquationRTF}
\end{equation}
\begin{equation}
    \frac{D^2 \xi^{\hat{i}}}{D\tau^2} = \Ddot{\xi}^{\hat{i}} = -\frac{2F_{0}}{\left(2 - 3\sqrt{F_{0}} t\right)^2} \xi^{\hat{i}}, \label{EquationATF}
\end{equation}
where $i = 2, 3$, corresponding to the angular components $\theta$ and $\phi$, respectively. Eqs. (\ref{EquationRTF}) and (\ref{EquationATF}) represent the tidal force equations for radially freely falling reference frames in the LTB metric for homogeneous dust collapse, with Eq. (\ref{EquationRTF}) describing the radial tidal force and Eq. (\ref{EquationATF}) describing the angular tidal force. These equations illustrate compression in the angular direction and stretching in the radial direction, collectively resulting in the phenomenon known as spaghettification \cite{madan2022tidal}.

Utilizing the tidal force components, we numerically demonstrate how the tidal force varies with physical radius ($R$) at different times, as depicted in Fig. (\ref{fig:1}) and Fig. (\ref{fig:2}) by the blue solid line.

From Eqs. (\ref{radial component of tidal force in homogeneous dust collapse}) and (\ref{angular component of tidal force in homogenenous dust collapse}), it is evident that the radial and angular components of the tidal force are equal. Given the homogeneous nature of the dust collapse, where density remains constant throughout the collapse, there is a consistent compressive effect. Therefore, in this scenario, the tidal force components remain independent of the comoving radius $r$ and vary solely with time $t$. Both the radial and angular tidal force values being negative indicate the presence of a compressive force.

\subsection{Tidal force effect in inhomogeneous dust collapse}
The LTB spacetime for the spherically symmetric inhomogeneous dust collapse is given by,
\begin{equation}
  \begin{aligned}
ds^{2} &= -dt^{2} + \frac{\left(2\sqrt{F_{0} - F_{3} r^{3}}-3F_{0} t+ 6 F_{3} r^{3} t\right)^{2}}{2^{4/3}(F_{0} - F_{3} r^3)\left(2-3\sqrt{F_{0} - F_{3} r^{3}}t\right)^{2/3}}dr^2 \\
&\quad + r^{2} \left(1-\frac{3}{2}\sqrt{F_{0} - F_{3} r^{3}}t\right)^{4/3}d\theta^{2} \\
&\quad + r^{2} \left(1-\frac{3}{2}\sqrt{F_{0} -F_{3} r^{3}}t\right)^{4/3} \sin^{2}\theta d\phi^2 \label{inhomo}.
\end{aligned}  
\end{equation}

Our investigation focuses on analyzing the tidal force equations embedded within the fabric of spacetime, specifically exploring the Lemaître-Tolman-Bondi (LTB) metric for the inhomogeneous dust collapse. For the Eq.~(\ref{inhomo}), the tetrad basis related to the freely falling reference frame can be expressed as \cite{2021EPJC...81..590L}:
\begin{align}
\hat{e}^{\mu}_{0} &= \left\{1,0,0,0\right\},\nonumber\\
\hat{e}^{\mu}_{1} &= \left\{0,\frac{2^{2/3}\sqrt{(F_{0} - F_{3} r^3)}\left(2-3\sqrt{F_{0} - F_{3} r^{3}}t\right)^{1/3}}{\left(2\sqrt{F_{0} - F_{3} r^{3}}-3F_{0} t+ 6 F_{3} r^{3} t\right)},0,0\right\}, \nonumber\\
\hat{e}^{\mu}_{2} &= \left\{0,0,\frac{1}{r \left(1-\frac{3}{2}\sqrt{F_{0} -F_{3} r^{3}}t\right)^{2/3}},0\right\},\nonumber\\
\hat{e}^{\mu}_{3} &= \left\{0,0,0,\frac{1}{r \left(1-\frac{3}{2}\sqrt{F_{0} - F_{3} r^{3}}t\right)^{2/3} \sin\theta}\right\}.\nonumber \label{tetradin}
\end{align}
Using the expression of Eq. (\ref{riemann tensor}), we can calculate the Riemann curvature tensor in terms of the tetrad basis as follows:\\
\begin{equation}{\label{radial component of tidal force in inhomogeneous dust collapse}}
R^{\hat{1}}_{\hat{0}\hat{0}\hat{1}} = \frac{2\left(-4\Psi\zeta + 12F_{0}^2\sqrt{\zeta}t - 2F_{3}r^3 \Sigma \sqrt{\zeta}t - 9\chi^2\zeta t^2\right)}{\left(2\sqrt{\zeta}-3\chi t\right)^2\left(2-3\sqrt{\zeta}t\right)^2}\, ,
\end{equation}
\begin{equation}{\label{angular component of tidal force in inhomogeneous dust collapse}}   R^{\hat{2}}_{\hat{0}\hat{0}\hat{2}}=R^{\hat{3}}_{\hat{0}\hat{0}\hat{3}}=-\frac{2\left(F_{0} - F_{3} r^3\right)}{\left(2-3\sqrt{F_{0} -F_{3}r^3}t\right)^2}\, ,
\end{equation}
where,  $\zeta=F_{0} - F_{3} r^3$;
 $\Psi = F_{0} - 4F_{3}r^3$;
 $\chi = F_{0} - 2F_{3}r^3$; $\Sigma=27F_{0} - 30F_{3}r^{3}$.
\begin{figure*}[]
\centering
\subfigure[]
{\includegraphics[width=55mm]{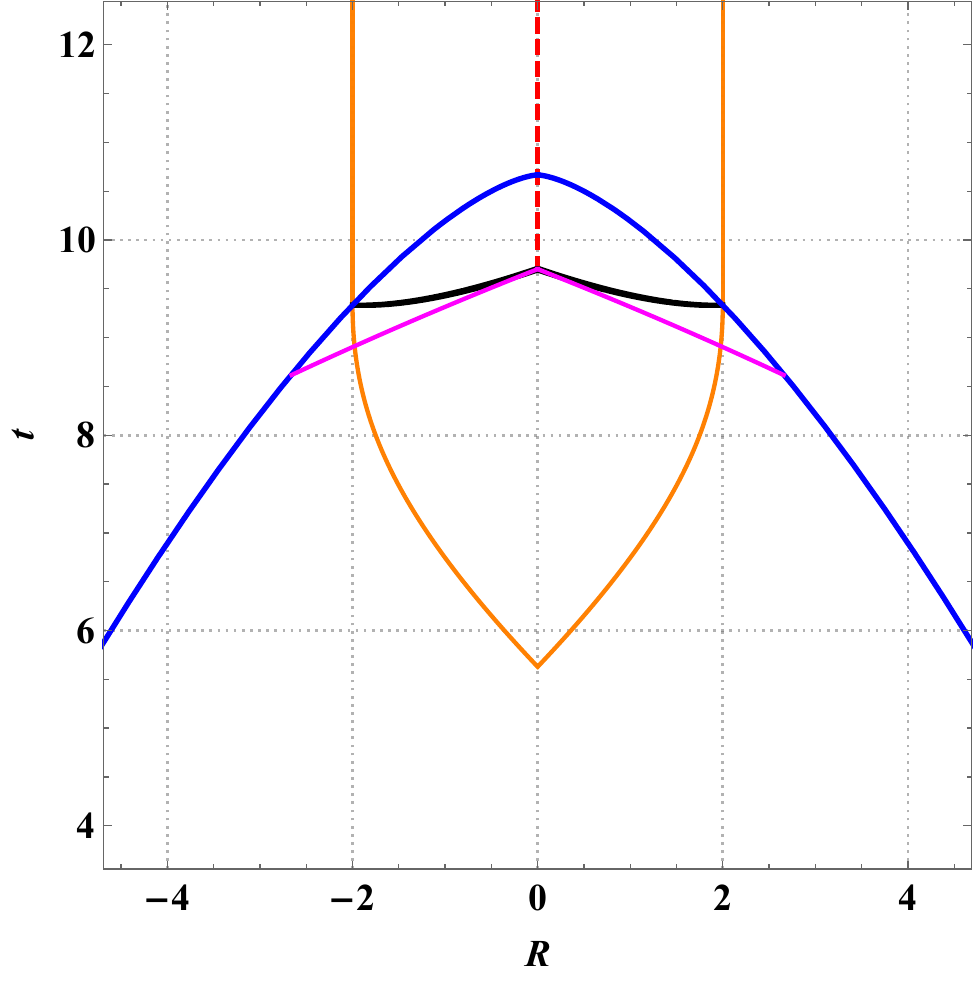}\label{ltbcollapseInHomo3}}
\hspace{0.5cm}
\subfigure[]
{\includegraphics[width=55mm]{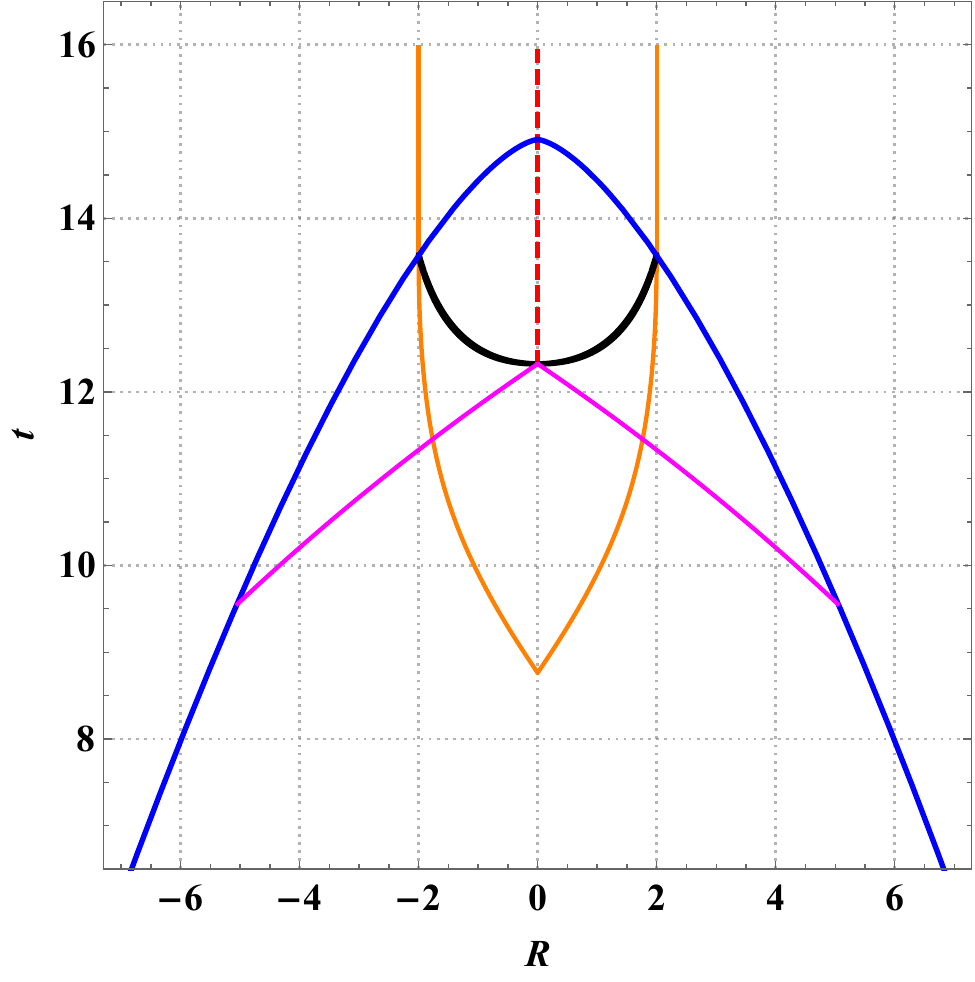}\label{ltbcollapseInHomo4}}
\hspace{0.5cm}
\subfigure[]
{\includegraphics[width=55mm]{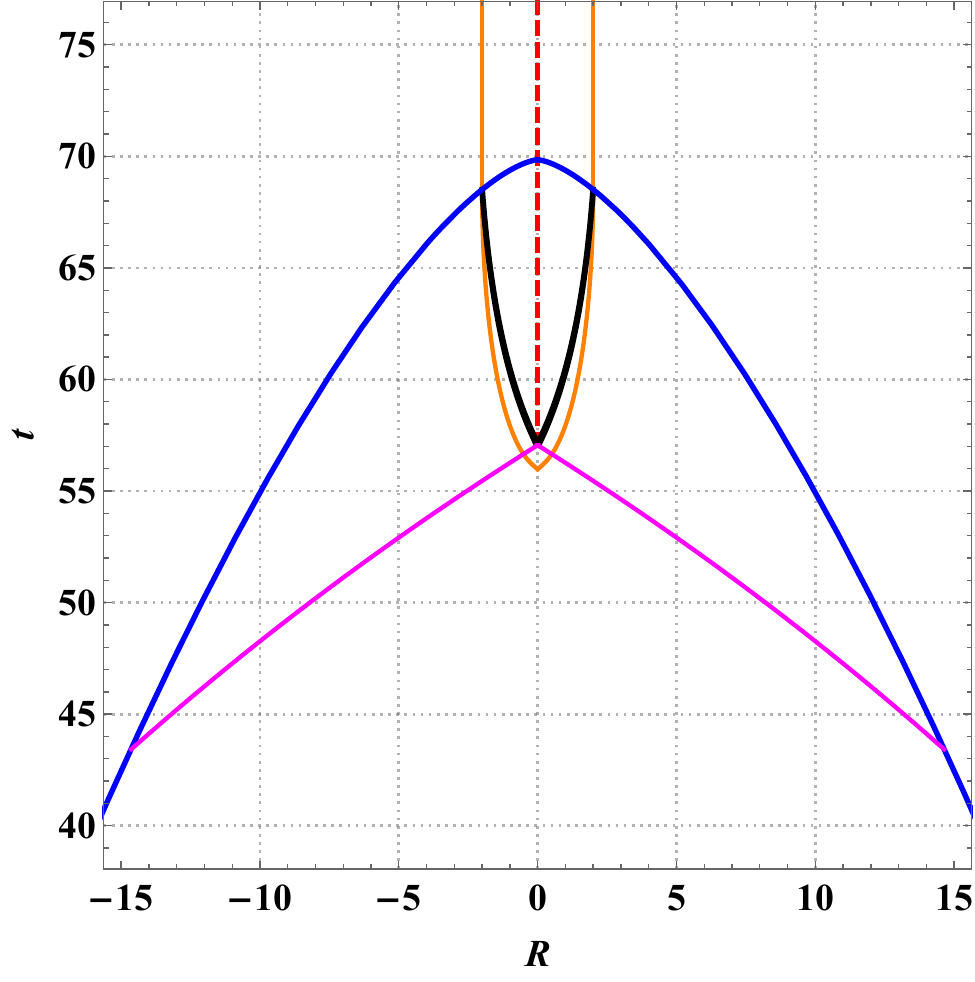}\label{penrose3}}
 \caption{This figure illustrates the gravitational collapse in the LTB metric with an inhomogeneous density profile, characterized by the mass function 
$ F(r) = M_{o}r^{3} - M_{3}r^{6} $. 
The solid yellow, blue, magenta and black lines represent the event horizon, the boundary of the collapsing cloud, the tidal critical boundary, and the apparent horizon, respectively. The numerical values for Fig.~(\ref{ltbcollapseInHomo3}) are 
$ M = 1 $, $ R_{b} = 8M $, and  $ \rho = 0.00927089 $. For Fig.~(\ref{ltbcollapseInHomo4}), the values are 
$ M = 1 $, $ R_{b} = 10M $, and 
$ \rho = 0.0032187 $. For Fig.(\ref{penrose3}), the values are $ M = 1 $, $ R_{b} = 28M $, and 
$ \rho = 0.000137 $.
}\label{InHomo}
\end{figure*}
In the same vein as the preceding homogeneous scenario, in Eq. (\ref{radial component of tidal force in inhomogeneous dust collapse}) and (\ref{angular component of tidal force in inhomogeneous dust collapse}), $R^{\hat{1}}_{\hat{0}\hat{0}\hat{1}}$ denotes the radial component of the tidal force, while $R^{\hat{2}}_{\hat{0}\hat{0}\hat{2}}$ and $R^{\hat{3}}_{\hat{0}\hat{0}\hat{3}}$ represent the angular components of the tidal force within the LTB metric for the inhomogeneous dust collapse. By substituting the values of the Riemann tensor into the geodesic deviation equation, we can derive the tidal force equation for the reference frame that is radially freely falling \cite{2021EPJC...81..590L} in LTB spacetime. The equation is given by:
\begin{equation}
        \frac{D^2 \xi^{\hat{1}}}{D\tau^2} = \Ddot{\xi}^{\hat{1}} = R^{\hat{1}}_{\hat{0}\hat{0}\hat{1}}\, \xi^{\hat{1}}\, ,\label{Radialinhomo}
\end{equation}
 \begin{equation}
    \frac{D^2\xi^{\hat{i}}}{D\tau^2}=\Ddot{\xi}^{\hat{i}}=R^{\hat{2}}_{\hat{0}\hat{0}\hat{2}}\, \xi^{\hat{i}}\,\label{anulartidal} ,
\end{equation}
where $i=2,3$ corresponds to the angular components $\theta$ and $\phi$, respectively. Equations (\ref{Radialinhomo}) and (\ref{anulartidal}) represent the tidal force equations for the radially freely falling reference frame in the context of inhomogeneous dust collapse. Using the expressions for tidal force components, we can generate plots illustrating the relationship between Tidal force and Physical radius ($R$) at various time intervals, as presented in Fig. (\ref{fig:2}).

One can calculate the time when radial tidal force vanishes i.e., when $R^{\hat{1}}_{\hat{0}\hat{0}\hat{1}} = 0$, and it comes out to be,
\begin{equation}
    t = \frac{2 \Psi}{3\chi \sqrt{\zeta}}\label{zerotidalradial}
\end{equation}
while in the angular case $R^{\hat{2}}_{\hat{0}\hat{0}\hat{2}} = 0$, gives a  condition $F_{0} = F_{3}r^3$.
We are also interested in the tidal force at the gravitational center at $r =0$. Therefore we can write the Eq.~(\ref{Radialinhomo}) and Eq.~(\ref{anulartidal}),
\begin{equation}
    \frac{D^2 \xi^{\hat{1}}}{D\tau^2} = \Ddot{\xi}^{\hat{1}} = \frac{2 F_{0}^2 \left(-4 + 12\sqrt{F_{0}}t - 9F_{0} t^2\right)}{\left(2\sqrt{F_{0}}-3F_{0} t\right)^2\left(2-3\sqrt{F_{0}}t\right)^2}\, \xi^{\hat{1}}
\end{equation}
\begin{equation}
    \frac{D^2\xi^{\hat{i}}}{D\tau^2}=\Ddot{\xi}^{\hat{i}}= -\frac{2 F_{0}}{\left(2-3\sqrt{F_{0}}t\right)^2}\, \xi^{\hat{i}}
\end{equation}

From the Eq.~(\ref{tidalhorizon}), for the Eq.~(\ref{Radialinhomo}) we can calculate,
\begin{widetext}
    \begin{eqnarray}
         -108 F_{3}^3 r^9 t^2 \left(-5 + \sqrt{F_{0} -  F_{3} r^3} t\right) + 72 F_{0}^2 \sqrt{F_{0} -  F_{3} r^3} t \left(4 + 3 F_{0} t^2\right) - 16 F_{0}^2 \left(4 + 27 F_{0} t^2\right) + \label{tsolve}\\ \nonumber 2  F_{3}^2 r^6 \left(-32 + 9 t \left(20 \sqrt{F_{0} -  F_{3} r^3} + 
       9 F_{0} t \left(-9 + 4 \sqrt{F_{0} -  F_{3} r^3} t\right)\right)\right) +\\ \nonumber F_{0}  F_{3} r^3 \left(128 -9 t \left(68 \sqrt{F_{0} -  F_{3} r^3} + 3 F_{0} t \left(-52 + 27 \sqrt{F_{0} -  F_{3} r^3} t \right) \right) \right) = 0,
    \end{eqnarray}
\end{widetext}
by solving (\ref{tsolve}) for variable $t$, we obtain three root solutions. By taking the condition that $t$ should not be greater the time of complete gravitational collapse that is, $t < 2/(3 \sqrt{F(R_{b})})$. This constrains give $t$ value as follows:
\begin{widetext}
    \begin{equation}
        t = \frac{\sqrt{2}}{3} \sqrt{\frac{128 F_{0}^4 - 432 F_{0}^3 F_{3} r^3 + 665 F_{0}^2 F_{3}^2 r^6 - 516 F_{0} F_{3}^3 r^9 + 164 F_{3}^4 r^{12} + X Y \sqrt{3} F_{1} r^3}{(F_{0} - F_{3} r^3) Z^2}}\label{troot},
    \end{equation}
    where,
    \begin{eqnarray}
        X &=& (-16 F_{0}^2 + 27 F_{0} F_{3} r^3 - 14 F_{3}^2 r^6), \nonumber\\
        Y &=& \sqrt{51 F_{0}^2 - 92 F_{0} F_{3} r^3 + 44 F_{3}^2 r^6}, \nonumber\\
        Z &=& 8 F_{0}^2 - 11 F_{0} F_{3} r^3 + 2 F_{3}^2 r^6.
    \end{eqnarray}
\end{widetext}
 From Eq.~(\ref{troot}) if $r =0$, we get $t = 2/(3\sqrt{F_{0}}$), i.e., maxima of tidal force is terminated and radial tidal force becomes infinite.
\begin{figure*}[ht!]
    \centering
\subfigure[]
{\includegraphics[width=8.0cm]{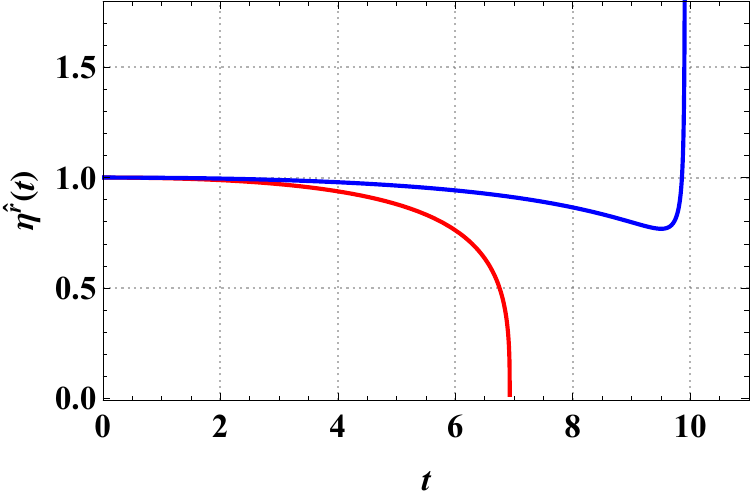}\label{jacobiradial}}
\hspace{0.5cm}
\subfigure[]
{\includegraphics[width=8.0cm]{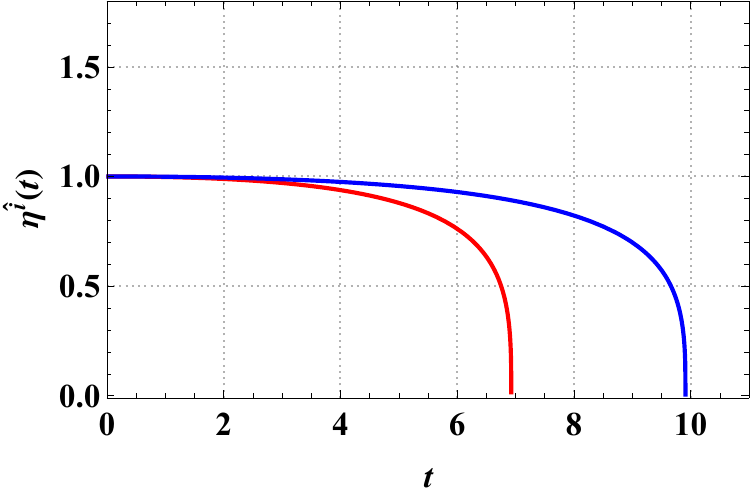}\label{jacobiangular}}
 \caption{In the above figures, the radial ($\eta^{\hat{r}}$) and angular ($\eta^{\hat{i}}$) components of the Jacobi field versus time $t$ for homogeneous (highlighted by red curve) and inhomogeneous (highlighted by blue curve) dust collapse are shown. Fig.~(\ref{jacobiradial}) illustrates how the radial components of the Jacobi field change with co-moving time $t$. Fig.~(\ref{jacobiangular}) demonstrates how the angular components of the Jacobi field change with co-moving time $t$ for a comoving radius of $r = 5M$. The initial conditions chosen are $\eta^{\hat{r}} = 1$, $\eta^{\hat{i}} = 1$, $\dot{\eta}^{\hat{r}} = 0$, and $\dot{\eta}^{\hat{i}} = 0$, where the overdot represents the derivative with respect to co-moving time $t$.}
 \label{figurejacobi}
\end{figure*}
\subsection{Comparative study of tidal force effect in the homogeneous and inhomogeneous dust collapse}{\label{sec IV}}
In this section, we present the results obtained from plotting graphs depicting tidal force's radial and angular components for both homogeneous and inhomogeneous dust collapse scenarios. Comparative graph plots illustrate the relationship between tidal force and physical radius ($R$) at various time intervals. For plotting we consider some values from the matching condition, which is given by, $F_{0} = 0.00925\qquad F_{1} = 5.9799\times 10^{-6}.$ Substituting the given values into the equation for singularity formation time yields $t_s=6.92$. Figs. (\ref{fig:1}) and (\ref{fig:2}) depicts the plots of radial and angular tidal force against physical radius ($R$) at different times. In these plots, negative radial and angular tidal force values signify compressive forces, while positive values indicate stretching forces.

In the homogeneous case, both radial and angular tidal forces consistently exhibit compression at different magnitudes across various time intervals. However, in the inhomogeneous case, the magnitude of the tidal force varies with radius at different time intervals. At $t \lessapprox 4.5$, the radial tidal force is compressive, while after $4.5$, compressive and stretching forces act according to the radial tidal distribution. As shown in Fig.~(\ref{fig:t=6}), at $t = 6.0$ within at $R = 1.5$ compressive nature is present while $1.5 \lessapprox R \lessapprox 2.1$ stretching force is present. As we can see from Fig.~(\ref{fig:t=6.5}), (\ref{fig:t=6.9}), (\ref{fig:t=6.92x}) that as time passes, compressive nature sharply increases near the coordinate center of collapsing body while at particular physical radius maxima of stretching force are present. After the formation of spacetime singularity that is $t > t_{s}(0)$, that is in Figs.~(\ref{fig:t=7.26 sec}) and (\ref{fig:t=7.35}) shows that radial tidal force is infinitely large near the spacetime singularity. In the situation of angular tidal force, a freely falling body towards the gravitational center is always compressive. In the homogeneous case, the compressive force within the collapsing body remains constant. However, in the inhomogeneous case, the angular tidal force is strongly compressive for small values of physical radius and weakly compressive for large values.

In conclusion, after the formation of the singularity, in the homogeneous case, the tidal deformation becomes infinite, rendering observation of this effect impossible. In contrast, in the inhomogeneous case, both the radial and angular tidal forces increase to infinity, indicating infinite radial stretching and angular compression, respectively. This culminates in the formation of a singularity that is infinitely tidally disrupted, known as the Ori strong singularity \cite{Ori:2000fi}.

\begin{figure*}[ht!]
    \centering
\subfigure[]
{\includegraphics[width=7.5cm]{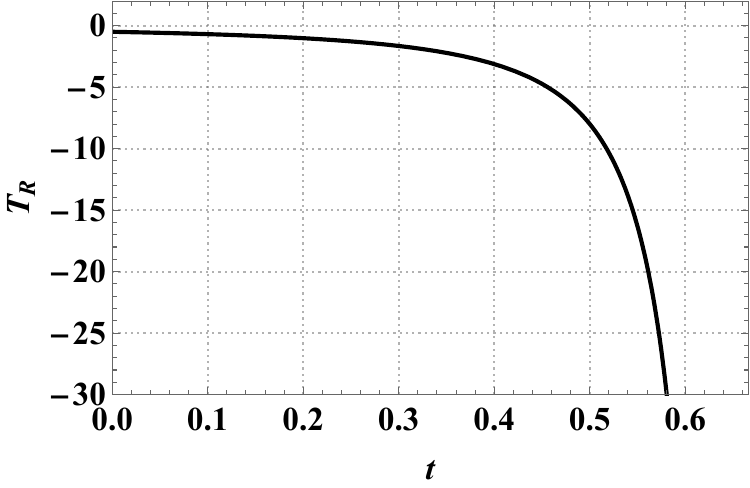}\label{GlobaltidalForceTrt}}
\hspace{0.5cm}
\subfigure[]
{\includegraphics[width=8.5cm]{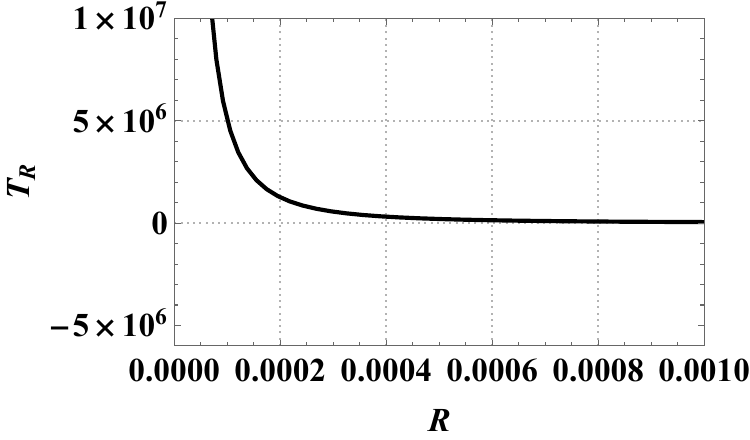}\label{GlobaltidalForceTrR}}
 \caption{Plots of radial tidal force ($T_R$) versus physical radius $(R)$ for inhomogeneous dust collapse with local visible singularity conditions, for $R_{b} = 0.2$ as shown in Eq.~(\ref{conditions}). Fig.~(\ref{GlobaltidalForceTrt}) shows radial tidal forces change with co-moving time, at central co-moving shell ($r=0$) and co-moving time $t$. Fig.~(\ref{GlobaltidalForceTrt}) shows radial tidal force changes with physical radius $R$ at time slices $t = t_{s}$.}
 \label{figG}
\end{figure*}

In Fig.~(\ref{GlobaltidalForceTrt}), we have shown the nature of tidal force at $r =0$ by varying time. However, for constant time slice by considering $t = t_{s}(0)$, we have shown in Fig.~(\ref{GlobaltidalForceTrR}) by taking $R_{b} = 0.2$, radial changes in tidal forces, as it is shown near the singularity, infinite radial tidal force. At the singularity, infinite compressive nature is present while at infinitesimally small vicinity, radial tidal forces have infinite positive value. This type of dual nature near the singularity affects the fluid around it. The interesting part is, in this case singularity is locally visible.

\subsection{Deformationally strong singularities}
The strength of spacetime curvature singularities can be used to categorize the nature of spacetime singularity. In \cite{Tipler:1977zza}, a definition of singularity strength based only on geometrical features of spacetime was put forth. That definition distinguishes between gravitationally weak and strong curvature singularities. Specifically, a spacetime singularity is referred to as strong if a volume (or area) element defined by linearly independent spacelike vorticity-free Jacobi fields propagating along any timelike (null) geodesic and orthogonal to its tangent vector vanishes at the singularity; if not, the singularity is referred to as weak. Strong curvature singularities require sufficient and necessary conditions, established in \cite{Clarke}. The definition mentioned above was later amended in \cite{Nolan:1999tw}, where each Jacobi field's behavior was taken into consideration. The updated definition states that if at least one Jacobi field vanishes or diverges at a spacetime singularity, the singularity is referred to as strong. When the volume element is still finite near the singularity despite certain Jacobi fields diverging and others vanishing, the singularity is referred to as strong \cite{Ori:2000fi}. By that definition, a spacetime singularity is referred to as \textit{deformationally strong} if at least one Jacobi field diverges and the volume element remains finite since other Jacobi fields vanish at the singularity. Otherwise, the volume element remains finite \cite{Abdolrahimi:2009dc}.

There is a necessary and sufficient condition that must be met in order to preserve the singularity's strength in the sense of a deformationally strong singularity. The criterion states that the following definition must satisfy for timelike  geodesic:\\

\textbf{Definition:} Let $\lambda(t)$ be a timelike geodesic with a proper time t along it. We shall say that $\lambda(t)$ terminates in a deformationally strong singularity at $t =0$ if at least one of the following two conditions hold: (I) $\lambda(t)$ terminates in a Tipler-strong singularity at $t=0$, or (II) there exists a Jacobi field $J(t)$ for which at least one PP tetrad component is unbounded at the limit $t \to 0$ \cite{Ori:2000fi}.

The norm of the 3-dimensional volume element of a synchronous frame, which is defined by 1-forms corresponding to the radial and angular Jacobi fields determined for the radial timelike geodesics as follows \cite{Abdolrahimi:2009dc},
\begin{equation}
    \lVert V(t) \rVert = | \eta^{\hat{r}} |  \prod_{n=2}^{3} | \eta^{\hat{i}} |.
\end{equation}
The following is an approximation of the norm of the volume element near the singularities based on the behavior of the Jacobi fields (for the given cases):\\
\textit{Case I: Homogeneous profile}
\begin{equation}
    \eta^{\hat{r}} = \eta^{\hat{i}} = C_{1} \left(2-3\sqrt{F_{0}}t\right)^{1/3} + C_{2} \left(2-3\sqrt{F_{0}}t\right)^{2/3}
\end{equation}
where $c_{1} = 2^{2/3}$ and $C_{2} = - 1/2^{2/3}$ for initial condition we choose $\eta^{\hat{r}}=1$, $\eta^{\hat{i}}=1$, $\dot{\eta}^{\hat{r}}=0$ and $\dot{\eta}^{\hat{i}} =0$. Here overdot represents derivative with respect to co-moving time $t$. For homogeneous case volume elements vanish at arbitrarily close to the singularity.\\
\textit{Case II: Inhomogeneous profile}
\begin{equation}
    \eta^{\hat{r}} \approx \frac{C_{1}}{(2 -3 \sqrt{\zeta} t)^{1/3}} - \frac{C_{2}(2-3 \sqrt{\zeta} t)^{4/3}}{5 \sqrt{\zeta}},
\end{equation}
where $C_{1} = 2^{7/3}/5$ and $C_{2} = - \sqrt{\zeta} / 2^{4/3}$ for given initial conditions. For the angular Jacobi field,
\begin{equation}
    \eta^{\hat{i}} = C_{3} (2 -3 \sqrt{\zeta} t)^{1/3} + \frac{C_{4} (2-3\sqrt{\zeta} t)^{2/3}}{\sqrt{\zeta}},
\end{equation}
as we can see near the singularity formation time angular Jacobi components $\eta^{\hat{i}}$ vanish. In the inhomogeneous case, volume elements for a given Jacobi field near the singularity formation time are written as:
\begin{widetext}
    \begin{align}
   \eta^{\hat{r}} \wedge \eta^{\hat{\theta}} \wedge \eta^{\hat{\phi}} \approx \frac{(2 - 3 \sqrt{\zeta} t)^{1/3}}{5 \zeta^{3/2}} \left(C_{4} (2 - 3 \sqrt{\zeta} t)^{1/3} + C_{3} \sqrt{\zeta}\right)^{2} \left(- C_{2} (2 - 3 \sqrt{\zeta} t)^{5/3} + 5 C_{1} \sqrt{\zeta}\right),
   \end{align}
\end{widetext}
These elements vanish as they approach the singularity, indicating that the singularity is Tipler strong. As stated in \cite{Magalhaes:2024smm}, Tipler's strong singularity is also a deformationally strong singularity, although the converse is not true.

The individual Jacobi fields for radial ($\eta^{\hat{r}}$) and angular ($\eta^{\hat{i}}$) components versus time $t$ for homogeneous and inhomogeneous dust collapse are shown in Fig.~(\ref{figurejacobi}). Fig.~(\ref{jacobiradial}) depicts how the radial components of the Jacobi field change with co-moving time $t$. Fig.~(\ref{jacobiangular}) illustrates how the angular components of the Jacobi field change with co-moving time $t$ for a comoving radius of $r = 5M$.

\section{CONCLUSION}{\label{sec3}}
From the equations of tidal force and the results of the plots, we can derive the following conclusions:

\begin{itemize}
\item The tidal forces in the static, spherically symmetric spacetime have been extensively studied. Not as much work is found in the dynamic scenario, though. In this case, we looked into the role of tidal force for both homogeneous and inhomogeneous density profiles in the LTB collapse. We also show the Jacobi field in the marginally bound case in LTB spacetime.

\item In the context of homogeneous dust fluid gravitational collapse, the tidal force effect remains independent of the comoving radius $r$ and varies over time. Our analysis reveals that in the homogeneous dust case, both radial and angular tidal forces consistently exhibit compression at different magnitudes across various time intervals, thus showing no variation with radius. At singularity formation time, tidal forces both angular and radial result in infinite compressive force.

\item In contrast, in the case of inhomogeneous dust collapse, the tidal force effect is radially distributed and time-dependent, allowing observation of both radial stretching and angular compression. Similarly as a homogeneous case, due to negative values of the angular tidal force, compressive forces predominate in the tangential part. Our analysis further indicates a change of sign in the radial tidal force, with positive values indicating stretching and negative values indicating compression. In the inhomogeneous case, the radial tidal force could give zero value as given by Eq.~(\ref{zerotidalradial}). After singularity formation, the radial tidal force increases to infinity in positive values, signifying infinite stretching, while the angular tidal force increases to infinity in negative values, signifying infinite compression. Consequently, infinite tidal deformation occurs in the inhomogeneous case, leading to the formation of the Ori strong singularity \cite{Ori:2000fi}. We have provided the best example that the visible singularity is the tidal strong singularity as shown in Fig.~(\ref{figG}).

In the case of the LTB collapse, the evolution of maximum stretching forces is given by Eq.~(\ref{tidalhorizon}). Due to the stretching forces that influence the fluid property, it could lead to more radiative at some radius. We consider a specific mass function ($F(r) = F_{0}r^{3} - F_{3}r^{6}$) to show the maximum stretching forces which are given by the Eq.~(\ref{troot}). Within the collapsing body, the maximum stretching forces could be evolved (represented by a solid magenta line) as shown in Fig.~(\ref{InHomo}). For easy to address we termed it a `critical tidal boundary'.

\end{itemize}
We tested the nature of spacetime singularity using spacetime geometry's tidal features. By investigating tidal deformation, we obtain insight into the changing causal structure of the spacetime geometry created during gravitational collapse. As previously stated, our findings shed light on how the tidal effect affects a freely falling object with low gravitational input as it approaches the center of a collapsing object. This understanding may have observational implications in our Universe's dynamic, ultra-high gravity regions.

\section{ACKNOWLEDGEMENT}
DD would like to acknowledge the support of the Atlantic Association for Research in the Mathematical Sciences (AARMS) for funding the work. AB Joshi acknowledges financial support from Ahmedabad University.


\begin{thebibliography}{99}
\bibitem{Joshi:2023ugm}
A.~B.~Joshi, K.~Mosani and P.~S.~Joshi,
\href{https://journals.aps.org/prd/abstract/10.1103/PhysRevD.109.064019}{Phys. Rev. D \textbf{109}, no.6, 064019 (2024)}

\bibitem{OppenheimerSnyder39} 
J. R. Oppenheimer and H. Snyder,
\href{https://journals.aps.org/pr/abstract/10.1103/PhysRev.56.455}{Phys. Rev. {\bf 56}, 455 (1939).}

\bibitem{Joshi:1993zg}
P.~S.~Joshi and I.~H.~Dwivedi,
\href{https://journals.aps.org/prd/abstract/10.1103/PhysRevD.47.5357}{Phys. Rev. D \textbf{47}, 5357-5369 (1993).}

\bibitem{gcsc} 
  Pankaj S. Joshi,
\href{https://doi.org/10.1017/CBO9780511536274}{Gravitational Collapse and Spacetime Singularities}

\bibitem{PhysRevLett.20.878}
A.~I.~Janis, E.~T.~Newman and J.~Winicour,
\href{https://journals.aps.org/prl/abstract/10.1103/PhysRevLett.20.878}{Phys. Rev. Lett. \textbf{20}, 878-880 (1968).}

\bibitem{perlick}
V Perlick
\href{https://iopscience.iop.org/article/10.1088/0264-9381/9/4/016}{Classical and Quantum Gravity, Volume 9, Number 4}

\bibitem{Joshi:2011zm}
P.~S.~Joshi, D.~Malafarina and R.~Narayan,
\href{https://iopscience.iop.org/article/10.1088/0264-9381/28/23/235018}{Class. Quant. Grav. \textbf{28}, 235018 (2011).}

\bibitem{Joshi:2020tlq}
A.~B.~Joshi, D.~Dey, P.~S.~Joshi and P.~Bambhaniya,
\href{https://journals.aps.org/prd/abstract/10.1103/PhysRevD.102.024022}{Phys. Rev. D \textbf{102}, no.2, 024022 (2020).}

\bibitem{Joshi:2013dva} 
  P. S. Joshi, D. Malafarina and R. Narayan,
\href{https://iopscience.iop.org/article/10.1088/0264-9381/31/1/015002/meta}{  Class. Quant. Grav.  {\bf 31}, 015002 (2014).}

\bibitem{Khodadi:2020gns}
M.~Khodadi and E.~N.~Saridakis,
\href{https://www.sciencedirect.com/science/article/abs/pii/S2212686421000662?via%3Dihub}{Phys. Dark Univ. \textbf{32}, 100835 (2021).}

\bibitem{Khodadi:2021gbc}
M.~Khodadi, G.~Lambiase and D.~F.~Mota,
\href{https://iopscience.iop.org/article/10.1088/1475-7516/2021/09/028}{JCAP \textbf{09}, 028 (2021).}

\bibitem{Khodadi:2022pqh}
M.~Khodadi and G.~Lambiase,
\href{https://journals.aps.org/prd/abstract/10.1103/PhysRevD.106.104050}{Phys. Rev. D \textbf{106}, no.10, 104050 (2022).}

\bibitem{KumarWalia:2022ddq}
R.~Kumar Walia,
\href{https://iopscience.iop.org/article/10.1088/1475-7516/2023/03/029}{JCAP \textbf{03}, 029 (2023).}

\bibitem{Tahelyani:2022uxw}
D.~Tahelyani, A.~B.~Joshi, D.~Dey and P.~S.~Joshi,
\href{https://journals.aps.org/prd/abstract/10.1103/PhysRevD.106.044036}{Phys. Rev. D \textbf{106}, no.4, 044036 (2022).}

\bibitem{Patra:2023epx}
S.~Patra, B.~R.~Majhi and S.~Das,
\href{https://arxiv.org/abs/2308.12839}{[arXiv:2308.12839 [astro-ph.HE]].}

\bibitem{Kovacs:2010xm} 
  Z. Kovacs and T. Harko,
\href{https://journals.aps.org/prd/abstract/10.1103/PhysRevD.82.124047}{Phys. Rev. D {\bf 82}, 124047 (2010).}

\bibitem{Joshi:2022azj}
A.~B.~Joshi, D.~Tahelyani, D.~Dey and P.~S.~Joshi,
\href{https://journals.aps.org/prd/abstract/10.1103/PhysRevD.108.104034}{Phys. Rev. D \textbf{108}, no.10, 104034 (2023).}

\bibitem{Kumar:2020ltt}
R.~Kumar and S.~G.~Ghosh,
\href{https://iopscience.iop.org/article/10.1088/1361-6382/abdd48}{Class. Quant. Grav. \textbf{38}, no.8, 8 (2021).}

\bibitem{Ghosh:2021txu}
R.~Ghosh and S.~Sarkar,
\href{https://journals.aps.org/prd/abstract/10.1103/PhysRevD.104.044019}{Phys. Rev. D \textbf{104}, no.4, 044019 (2021).}

\bibitem{Shaikh:2022ivr}
R.~Shaikh,
\href{https://academic.oup.com/mnras/article-abstract/523/1/375/7157137?redirectedFrom=fulltext}{Mon. Not. Roy. Astron. Soc. \textbf{523}, no.1, 375-384 (2023).}

\bibitem{Bambhaniya:2019pbr}
P.~Bambhaniya, A.~B.~Joshi, D.~Dey and P.~S.~Joshi,
\href{https://journals.aps.org/prd/abstract/10.1103/PhysRevD.100.124020}{Phys. Rev. D \textbf{100}, no.12, 124020 (2019).}

\bibitem{Battista:2022krl}
E.~Battista and G.~Esposito,
\href{https://link.springer.com/article/10.1140/epjc/s10052-022-11070-w}{Eur. Phys. J. C \textbf{82}, no.12, 1088 (2022).}

\bibitem{madan2022tidal}
S.~Madan and P.~Bambhaniya,
\href{https://arxiv.org/pdf/2201.13163.pdf}{arXiv:2201.13163 [gr-qc](2022)}

\bibitem[Lima Junior et al.(2022)]{zhang2018tidal} Lima Junior, H.~C.~D., Corr{\^e}a, M.~M., Macedo, C.~F.~B., et al.\ 2022, European Physical Journal C, 82, 479. 
\href{https://arxiv.org/pdf/2205.13569.pdf}{doi:10.1140/epjc/s10052-022-10410-0}

\bibitem{Toshmatov:2023anz}
B.~Toshmatov and B.~Ahmedov,
\href{https://journals.aps.org/prd/abstract/10.1103/PhysRevD.108.084035}{Phys. Rev. D \textbf{108}, no.8, 084035 (2023).}

\bibitem{Vandeev_2021}
V.~P.~Vandeev and A.~N.~Semenova,
Eur. Phys. J. C \textbf{81}, no.7, 610 (2021)
\href{https://doi.org/10.1140/epjc/s10052-021-09427-8}{doi:10.1140/epjc/s10052-021-09427-8}

\bibitem{Arora:2023ijd}
D.~Arora, N.~U.~Molla, H.~Chaudhary, U.~Debnath, F.~Atamurotov and G.~Mustafa,
\href{https://arxiv.org/abs/2308.13901}{arXiv:2308.13901 [gr-qc].}

\bibitem{Arora:2023ltv}
D.~Arora, P.~Bambhaniya, D.~Dey and P.~S.~Joshi,
\href{https://arxiv.org/abs/2305.08082#}{arXiv:2305.08082 [gr-qc].}
[arXiv:2305.08082 [gr-qc]].

\bibitem{Rntidal}
L.~C.~B.~Crispino, A.~Higuchi, L.~A.~Oliveira and E.~S.~de Oliveira,
\href{https://link.springer.com/article/10.1140/epjc/s10052-016-3972-5}{Eur. Phys. J. C \textbf{76}, no.3, 168 (2016).}

\bibitem[Liu et al.(2022)]{Li2017} Liu, J., Chen, S., \& Jing, J.\ 2022, Chinese Physics C, 46, 105104. 
\href{https://arxiv.org/pdf/2203.14039.pdf}{{doi:10.1088/1674-1137/ac7856}}

\bibitem[Li et al.(2021)]{2021EPJC...81..590L} Li, J., Chen, S., \& Jing, J.\ 2021, European Physical Journal C, 81, 590.
\href{https://arxiv.org/pdf/2105.01267.pdf}{ doi:10.1140/epjc/s10052-021-09400-5}

\bibitem{wald}
R.~M.~Wald,
\textit{General Relativity},
(\href{https://press.uchicago.edu/ucp/books/book/chicago/G/bo5952261.html}{Chicago University Press, 1984}).

\bibitem{hawking} S. W. Hawking and G. F. R. Ellis, \textit{The large scale structure of spacetime}, (\href{https://www.cambridge.org/core/books/large-scale-structure-of-spacetime/1E6B961EC9878EDDBBD6AC0AF031CC93}{Cambridge University Press, 1973}).

\bibitem{Tipler:1977zza}
F.~J.~Tipler,
\href{https://www.sciencedirect.com/science/article/abs/pii/0375960177905084?via%3Dihub}{Phys. Lett. A \textbf{64}, 8-10 (1977).}

\bibitem{Clarke}
C.J.S. Clarke, A. Królak,
\href{https://www.sciencedirect.com/science/article/abs/pii/0393044085900129?via%3Dihub}{Journal of Geometry and Physics \textbf{2}, 127 (1986).}

\bibitem{Newman:1985gt}
R.~P.~A.~C.~Newman,
\href{https://iopscience.iop.org/article/10.1088/0264-9381/3/4/007 }{Class. Quant. Grav. \textbf{3}, 527-539 (1986).}

\bibitem{Waugh:1988ud}
B.~Waugh and K.~Lake,
\href{https://journals.aps.org/prd/abstract/10.1103/PhysRevD.38.1315}{Phys. Rev. D \textbf{38}, 1315-1316 (1988).}

\bibitem{Nolan:1999tw}
B.~C.~Nolan,
\href{https://journals.aps.org/prd/abstract/10.1103/PhysRevD.60.024014}{Phys. Rev. D \textbf{60}, 024014 (1999).}

\bibitem{Ori:2000fi}
A.~Ori,
\href{https://journals.aps.org/prd/abstract/10.1103/PhysRevD.61.064016}{Phys. Rev. D \textbf{61}, 064016 (2000).}

\bibitem{refn1} 
  E.~Poisson,
\href{https://www.cambridge.org/core/books/relativists-toolkit/DA7ED68B971708A0F782257F948981E7}{Cambridge University Press. (2004) doi:10.1017/CBO9780511606601 
}

\bibitem{Israel:1966rt}
W.~Israel,
\href{https://link.springer.com/article/10.1007/BF02710419}{Nuovo Cim. B \textbf{44S10}, 1 (1966).}; 
[erratum: \href{https://link.springer.com/article/10.1007/BF02712210}{Nuovo Cim. B \textbf{48}, 463 (1967).}]

\bibitem{Joshi:2024kqd}
A.~B.~Joshi,
\href{https://arxiv.org/abs/2402.15156}{arXiv:2402.15156 [gr-qc].}

\bibitem{chandrasekhar1983mathematical}
Chandrasekhar, S.
\emph{The Mathematical Theory of Black Holes}. Clarendon Press, 1983.

\bibitem{Abdolrahimi:2009dc}
S.~Abdolrahimi and A.~A.~Shoom,
\href{https://journals.aps.org/prd/abstract/10.1103/PhysRevD.81.024035}{Phys. Rev. D \textbf{81}, 024035 (2010).}

\bibitem{Magalhaes:2024smm}
R.~B.~Magalh\~aes, G.~P.~Ribeiro, H.~C.~D.~Lima, Junior, G.~J.~Olmo and L.~C.~B.~Crispino,
\href{https://iopscience.iop.org/article/10.1088/1475-7516/2024/05/114}{JCAP \textbf{05}, 114 (2024).}

\end{thebibliography}
\end{document}